\documentclass[sigconf]{acmart}

\usepackage{booktabs}
\usepackage{multirow}
\usepackage{tabularx}
\usepackage{array}
\newcolumntype{L}[1]{>{\raggedright\arraybackslash}m{#1}} % 세로 가운데 + 왼쪽 정렬
\newcolumntype{R}[1]{>{\raggedright\arraybackslash}p{#1}} % 일반 단락형 + 왼쪽 정렬

\usepackage{enumitem}
\usepackage{ragged2e}
\usepackage{tabularx}
\usepackage{graphicx}
\usepackage{todonotes}
\usepackage{booktabs}
\usepackage{enumitem}

\usepackage{soul}    % defines \st (strikeout), \hl, \ul
\usepackage{xcolor}

\let \oldst \st     %creates \oldst to do the role of \st so that we can redefine \st's function
\let \oldtextcolor \textcolor  %creates \oldtextcolor to do the role of \textcolor so that we can redefine \textcolor's function

\renewcommand{\st}[1]{\oldtextcolor{red}{\oldst{#1}}} %re-define \st command to print red strikethrough text

\AtBeginDocument{%
  \providecommand\BibTeX{{%
    \normalfont B\kern-0.5em{\scshape i\kern-0.25em b}\kern-0.8em\TeX}}}

\setcopyright{acmlicensed}
\copyrightyear{2026}
\acmYear{2026}
\setcopyright{cc}
\setcctype{by}
\acmConference[CHI '26]{Proceedings of the 2026 CHI Conference on Human Factors in Computing Systems}{April 13--17, 2026}{Barcelona, Spain}
\acmBooktitle{Proceedings of the 2026 CHI Conference on Human Factors in Computing Systems (CHI '26), April 13--17, 2026, Barcelona, Spain}
\acmPrice{}
\acmDOI{10.1145/3772318.3790429}
\acmISBN{979-8-4007-2278-3/2026/04}

\begin{document}

\title{The Values of Value in AI Adoption: Rethinking Efficiency in UX Designers' Workplaces}
% \title{Contesting Efficiency in AI Adoption: Negotiating the Meanings of "Value" in AI Adoption in UX Designers' Workplaces}

\author{Inha Cha}
\email{icha9@gatech.edu}
\affiliation{%
  \institution{Georgia Institute of Technology}
  \city{Atlanta}
  \state{GA}
  \country{USA}
  \postcode{30308}
}

\author{Catherine Wieczorek}
\email{cwieczor3@gatech.edu}
\affiliation{%
  \institution{Georgia Institute of Technology}
  \city{Atlanta}
  \state{GA}
  \country{USA}
  \postcode{30308}
}

\author{Richmond Y. Wong}
\email{rwong34@gatech.edu}
\affiliation{%
  \institution{Georgia Institute of Technology}
  \city{Atlanta}
  \state{GA}
  \country{USA}
  \postcode{30308}
}

% Negotiating the Values of Value in AI Adoption: Examining the Role of Efficiency in UX Workplaces

% The (Social) Value of (Economic) Value: Negotiating the Values of AI Adoption in Designers' Workplaces

% The (Social) Value of (Economic) Value: Re-examining the Role of Efficiency in Designers' Workplaces

% Negotiating the Efficiency of AI Adoption in Designers' Workplaces 

% Reconfiguring the Value of AI Adoption in Designers' Work

% Contesting Efficiency in AI Adoption: Designer and Organizational Perspectives on the Value of AI Adoption

% \affiliation{%

% \affiliation{%

% \affiliation{%

\renewcommand{\shortauthors}{Cha et al.}

\begin{abstract}
% Organizations increasingly frame AI adoption as improving productivity and efficiency. However, organizational perspectives on the (economic) value of AI may be at odds with workers' perspectives on the (economic and social) value of AI. Through design workshops with 15 designers, we present findings about the dynamics of AI adoption at designers' organizations, showing how different tensions arise at individual, team, and organizational scales. Across these scales, tensions arise in how the economic of social "value" of AI is considered, the ways that AI may reconfigure work responsibilities, and the potential uneven benefits and harms of these changes. We contribute _________________. 

Although organizations increasingly position AI adoption as a pathway to competitiveness and innovation, organizations' perspectives on productivity and efficiency often clash with workers' perspectives on AI's economic and social value. Through design workshops with 15 UX designers, we examine how AI adoption unfolds across individual, team, and organizational scales. At the individual level, designers weighed efficiency, skill development, and professional worth. At the team level, they negotiated collaboration, responsibility, and rigor. At the organizational level, adoption was shaped by compliance requirements and organizational norms. Across these scales, discourses of efficiency carried social and ethical dimensions of responsibility, trust, and autonomy. We view adoption as a site where roles, relationships, and power are reconfigured. We argue that AI adoption should be understood as a process of negotiating values, and call for future work examining how AI systems redistribute responsibility among team members, while understanding how such shifts could strengthen worker agency.

\end{abstract}

\begin{CCSXML}
<ccs2012>
   <concept>
       <concept_id>10003120.10003121.10011748</concept_id>
       <concept_desc>Human-centered computing~Empirical studies in HCI</concept_desc>
       <concept_significance>500</concept_significance>
       </concept>
 </ccs2012>
\end{CCSXML}

\ccsdesc[500]{Human-centered computing~Empirical studies in HCI}

\keywords{AI, AI Adoption, Workplace, Tech Practitioners, UX Practitioners, UX Practices}

% \received{September 2025}
% \received[revised]{November 2025}
% \received[accepted]{January 2026}

\maketitle
%%%%%%Markup%%%%
% \section{Introduction}
% \input{1_Introduction_markup}
%  \section{Related Works}
%  \input{2_RelatedWorks_markup}
%   \section{Method}
%  \input{3_Methods_markup}
% \section{Findings: Negotiating the Values of AI Adoption}
% \input{4_Findings_markup}
% \section{Discussion}
% \input{5_Discussion_markup}
% \section{Conclusion}
% \input{6_Conclusion_markup}

%%%%%%clean%%%%%%%%%%

\section{Introduction}
Organizations increasingly frame emerging technologies as critical opportunities for competitiveness and innovation \cite{spanjol2018successive, yee2025superagency}. Artificial Intelligence (AI), in particular, has been positioned both as a new industrial revolution \cite{verganti2020innovation, kassa2025impact} and as a central force for innovation \cite{mariani2023types}. Design and User Experience (UX) domains have become prominent sites of AI experimentation \cite{shin2025what, 10.1145/3706598.3713273}, where work is often viewed as cost- and labor-intensive \cite{shin24DIS}. AI tools are now being deployed to enhance productivity and expand creative possibilities, from synthetic persona generation \cite{shin24DIS} to visual ideation \cite{10.1145/3613904.3642847} to automated qualitative data analysis \cite{liCHI24}.

Recent HCI scholarship has begun to move beyond evaluating the task-level effectiveness and efficiency of these AI tools, instead calling for studying the social and organizational dynamics of how AI is adopted and used in UX practices (e.g., \cite{cha25chi, 10.1145/3643834.3660703}). This shift resonates with a broader turn in HCI from a user-centered paradigm toward an adoption-centered perspective \cite{10.1145/3025453.3025742, 10.1145/2702123.2702412}. While traditional HCI emphasized usability and user needs \cite{cooper1995representing}, adoption-centered research has shown that the long-term viability of technology depends on multiple stakeholders who shape adoption decisions beyond end users. Prior work has largely examined how research prototypes persist beyond the lab \cite{10.1145/2702123.2702412}, but has paid less attention to how social values are surfaced and negotiated in adoption processes.

% \st{To situate this shift in focus, it is useful to recall how HCI has historically conceptualized efficiency as a central evaluative lens. Efficiency has long guided HCI, from early workplace studies and human factors research that emphasized nondiscretionary interfaces} \cite{grudin2012moving}
% \st{, to usability testing that measured efficiency, effectiveness, and satisfaction} \cite{myers1998brief, carroll2001evolution}
% \st{. Even as “second and third wave” HCI expanded beyond work and rationality} \cite{10.1145/2804405}
% \st{, efficiency has persisted as a guiding orientation. This persistence invites us to rethink what efficiency means. Efficiency itself has always been contested: Greenbaum }\cite{greenbaum1979name} 
% \st{distinguishes between quantitative efficiency (measurable output) and qualitative efficiency (minimizing resistance). Productivity can be shaped by both, and gains in one may undermine the other, a tension that also surfaces in the difference between individual and team productivity. This distinction highlights that efficiency is a concept whose meaning \emph{shifts} depending on context and perspective. Contemporary management research similarly equates productivity with tighter managerial control over workers, which provokes pushback from workers and underscores the need for alternative models that privilege their autonomy and effectiveness} \cite{omidi2023labor}
% \st{. These debates indicate that the meaning of efficiency ultimately depends on who defines it and who benefits.} 

We build on HCI and CSCW research on workplace technologies, which shows that adoption cannot be treated as a purely technical matter. Previous research has noted that workplace systems are often designed to reflect the priorities of upper-level management rather than the needs and perspectives of workers \cite{10.1145/3715070.3748296, beaudry2005changes, 10.1145/3715070.3747335, 10.1145/3491101.3516386}. For instance, Greenbaum \cite{10.1145/240080.240259} highlights how CSCW systems can equip management to shift tasks onto a workforce that is both specialized and low-cost, prompting questions about whose needs these systems truly advance.
% Feminist and postcolonial perspectives further demonstrate that technological systems embed particular vantage points and distribute benefits and burdens unevenly \cite{10.1145/3686899, 10.1145/3744169.3744179}. 
Acknowledging that feminist and postcolonial perspectives further demonstrate that technology particular vantage points and distribute benefits and burdens unevenly~\cite{10.1145/3686899, 10.1145/3744169.3744179}, we turn attention to how teams manage these tensions. 
From this perspective, adoption becomes a moment in which efficiency, optimization, and value are not neutral, but situated—advancing some actors and their practices while constraining others. Such tensions indicate that adopting AI in UX work is not only a matter of technical integration but a site where different beliefs about what constitutes “value” come into friction.
Extending these conversations, we shift attention from which AI tools are adopted to how adoption itself is deliberated within design work. Rather than treating adoption as a simple take-it-or-leave-it decision \cite{10.1145/3744169.3744170}, we frame adoption as a situated process of using, not using, and selectively using AI tools shaped by questions of control, labor, and organizational resources. 
% \st{Therefore, w}
We conducted design workshops and individual follow-up interviews with 15 designers, examining how UX professionals deliberate on AI adoption in their everyday work. Specifically, we ask: 
% \st{What do UX designers encounter and experience in the process of AI adoption at the individual, team, and organizational levels?}
how do UX professionals experience and negotiate AI adoption in their everyday work practices?

In our findings, AI adoption unfolds as an ongoing process of aligning stakeholders, reconciling tensions, and reconfiguring practices across scales. At the individual level, designers evaluated AI in terms of efficiency, skill development, and long-term professional worth. At the team level, they grappled with how AI reshaped collaboration, responsibility, and rigor. At the organizational level, adoption was mediated by formal structures, compliance requirements, and broader industry forces, with AI integration reflecting both designers’ agency and the cultural and power dynamics of their organizations. Our findings show that the central issue is not whether AI ``adds value,'' but ``what does value actually mean.'' Therefore, these perspectives show that AI adoption is not a discrete event or a technical decision, but an ongoing process of negotiating values, aligning stakeholders, and reconfiguring practices across individual, team, and organizational contexts.

To understand how differing notions of value surface when AI tools enter UX practice, we draw on Graeber’s distinction \cite{Graeber_2001} between \textit{value} as economic worth and \textit{values} as moral and social commitments.\footnote{As David Graeber observed in tracing the term’s intellectual genealogy, ``value'' has carried at least three overlapping senses in social theory: (1) ideals of what is good or desirable (``values''), (2) economic worth, often measured through exchange (``value''), and (3) linguistic value, or the meaningful differences that structure systems of signs.} Describing value as the ``false coin of our own dreams,” he reminds us value cannot be reduced to metrics of productivity or price but is always bound up with the broader ideals and social commitments through which people imagine what matters in their lives.
Building on this view, we treat value in this study not as singular or self-evident, but as contested and situated. When AI tools enter UX practice, multiple—and sometimes competing—interpretations of value emerge, ranging from quantifiable productivity metrics \cite{10.1145/2832117} to relational and organizational matters such as collaboration, autonomy, and workplace culture.

Our discussion therefore centers on how AI adoption is less about technical efficiency than about their relational meaning: the roles tools assume within teams, the meanings attached to them, and their effects on workflows, relationships, and decision-making. We show how discourses of efficiency often carried social dimensions (e.g., responsibility, trust, autonomy) revealing adoption as a process of reconfiguring who does what, who decides, and whose contributions count. For future research, we suggest how AI shifts responsibility and reshapes organizational politics, producing uncertainty and identity struggles for workers \cite{Mulligan_Nissenbaum_2020}. AI adoption also alters relational dynamics within design work, not only in how teams coordinate, but in how workers care for, trust, and support one another. These forms of relational labor become entangled with automation, delegation, and organizational expectations, sometimes strengthening interdependence and at other times exposing tensions and invisible work. We suggest that future research examine how AI reconfigures relational labor and reshapes how relationships are maintained, sustained, and valued at work.
We conclude by calling for future research that develops mechanisms to strengthen worker agency in deciding if, when, and how AI should be integrated.

\section{Related Works}
\subsection{Efficiency as Contested Value in Workplaces}

\subsubsection{Workplace Technologies and Efficiency}
For centuries, artisans relied on specialized tools, but systematic efforts to improve work efficiency began in the early 20th century with Taylor’s scientific management \cite{uddin2015evolution} and time-and-motion studies \cite{grudin2005three}. World Wars I and II accelerated efficiency and human factors research, as complex military equipment revealed the limits of human capabilities and spurred ergonomic design improvements \cite{waterson2011world}. 
% \st{This wartime research prioritized reducing training requirements while critically focusing on error elimination and accelerating skilled task performance} \cite{roscoe1997adolescence}\st{. These foundational developments established modern human factors principles, recognizing both computing's transformative potential and the centrality of behavioral requirements in design and training initially within nondiscretionary human-machine interfaces} \cite{grudin2012moving}.
% \st{Throughout its evolution, HCI has retained a strong efficiency orientation, progressing from task-focused experiments and psychology-based studies to} 
These developments established modern human factors principles and reinforced an efficiency-oriented paradigm early HCI, which moved from task-focused experiments to structured usability testing that measured efficiency, effectiveness, and satisfaction \cite{myers1998brief, carroll2001evolution}. As Lin et al. point out, a focus on ‘use’ has continued to shape discussions in the HCI field, governing its methods and design decisions in ways that reinforce the promise of progress and productivity \cite{10.1145/3411764.3445237}. 
% \st{Over time, the field expanded to encompass usability principles, emotional design, and broader user experience frameworks. This trajectory crystallized into an engineering-driven paradigm that expected scientifically grounded methods to yield predictable, quantifiable outcomes. At the same time, HCI became increasingly interdisciplinary, drawing on psychology, computer science, ergonomics, and the social sciences. With computing becoming smaller, cheaper, and ubiquitous, both user populations and application domains expanded, producing far-reaching societal implications}
As computing became smaller, cheaper, and ubiquitous, user populations and application domains expanded, and efficiency emerged not only as a technical attribute of systems but also as a value negotiated among diverse actors and institutional constraints \cite{blythe2004funology, 10.1145/1357054.1357156, dix2017human}. 

\subsubsection{Efficiency as Relational Value in Teams and Organizations}

% \st{The growing diversity of contexts and implications in HCI makes it clear that efficiency cannot be treated as a singular, uncontested idea. Rather, it is a multifaceted and often disputed concept.} 
Across HCI and ICTD, scholars have shown that seemingly neutral ideals like efficiency and automation are entangled with colonial histories \cite{irani2010postcolonial}, gendered forms of invisible labor \cite{rosner2018critical}, and the shifting demands placed on care within sociotechnical infrastructures \cite{karusala2023unsettling}. While our aim is not to advance a full postcolonial or care-ethical critique of AI and efficiency, we take seriously D’Ignazio and Klein’s call to interrogate how notions of ``efficiency'' reflect particular values and power relations rather than universal technical truths \cite{d2023data}. From this perspective, efficiency becomes a situated concept that workers grapple with in their everyday work. Thus, in this paper we re-interrogate ``efficiency’’ through longer lineages of the concept within HCI, with specific attention to how it is relationally produced and contested in organizational settings.

Greenbaum’s seminal work \textit{In the Name of Efficiency} \cite{greenbaum1979name}, building on Gordon’s distinction between quantitative efficiency (increasing measurable output) and qualitative efficiency (minimizing worker resistance), highlights the tension between these competing dimensions. Attempts to improve one often undermine the other, revealing efficiency not as a neutral metric but as a site of sociopolitical negotiation. Greenbaum further reframes this duality in terms of individual productivity versus team productivity, suggesting that what is often at stake is less about “two kinds of efficiency” than about two kinds of productivity: efficiency and effectiveness. Management science often pursues quantitative productivity through heightened control, such as dividing labor into narrower specialties, restricting worker discretion, and privileging quantifiable activities, usually at the cost of increased resistance \cite{omidi2023labor}. By contrast, workers seek greater agency as artisans, aspiring to apply judgment and experience in ways that escape strict measurement \cite{greenbaum1979name}. This tension is vividly illustrated in the case of free and open-source software adoption in programming, where programmers resisted the costly enterprise tools procured by management and instead embraced freely available alternatives without approval, thereby reclaiming control over the means of production. As Greenbaum observes, “efficient work activities can take place without the management ideology of social control,” pointing toward alternative forms of work organization that foreground effectiveness and worker autonomy over managerial efficiency.

These dynamics resonate with critical labor process theory, which has been critiqued for underplaying the potential for worker resistance, overlooking how workers actively negotiate, resist, and reshape managerial control \cite{smith2015continuity}. They also align with HCI’s long-standing interest in workplace democracy and technology adoption (e.g., Scandinavian participatory tradition \cite{gregory2003scandinavian, muller1993participatory, schuler1993participatory, goransdotter2022designing}, dissensus-based workplace technology design \cite{10.1145/3641001}). 
%These historical tensions anticipate contemporary HCI debates, which extend beyond individual productivity to encompass collective dynamics, organizational values, and situated contexts. Such concerns directly inform our inquiry into how efficiency is conceptualized and contested within current AI and UX practices, across their varied positionalities.
In context, efficiency becomes a site of conflict across positionalities within organizational settings, as stakeholders experience and define what counts as ``efficient'' in different ways. These historical and conceptual tensions surface in our findings on how efficiency is conceptualized, produced, and contested in contemporary AI and UX practices across these varied positionalities.

\subsection{AI and Social Dynamics in UX Design Practices}
AI has been described as both a new industrial revolution \cite{verganti2020innovation, kassa2025impact} and a major driver of innovation \cite{mariani2023types}. HCI researchers have explored AI across domains such as healthcare \cite{wieczorek2025architecting}, education \cite{ding2025considering}, journalism \cite{10.1145/3715275.3732198}, and design \cite{10.1145/3377325.3377522}. Within design, UX has become a prominent site for AI integration \cite{10.1145/3643834.3660720}. For example, Choi et al. \cite{10.1145/3613904.3642794, 10.1145/3706599.3720189} developed generative systems that help designers extract visual elements from reference images using keywords to support ideation, while Son et al. \cite{10.1145/3613904.3642847} incorporated generative models into visual search to aid early-stage exploration. AI has also been applied to UX workflows such as persona creation \cite{shin24DIS}, journey mapping \cite{10.1145/3706598.3713479}, and prototyping \cite{li2021ml}. However, much of this research emphasizes user experience and artifact quality, with fewer studies addressing user-centric benefits. Recent work \cite{10.1145/3698061.3726924} calls for examining how such technologies improve not only creative outputs but also the quality of life of those engaged in creative work.

Recent work emphasizes the nonlinear nature of design when AI tools are introduced \cite{10.1145/3706599.3720189}. Cha et al. \cite{cha25chi} argue that narrow evaluation methods overlook how AI integrates into broader workflows. Furthermore, Wang et al. \cite{10.1145/3643834.3660703} examined collaboration with AI, showing that it can reduce friction by accelerating visualization tasks or enabling non-specialists to contribute. At the same time, AI introduced new tensions, as outputs sometimes surpassed those of collaborators, raising conflict when suggestions were ignored.

In UX practice, collaboration within and across teams is both crucial and complex, as interpersonal communication is inherently relational, dynamic, and context-sensitive \cite{liCHI24, 10.1145/3643834.3660703, 10.1145/3643834.3660720}. 
The introduction of AI tools adds another layer of complexity, shaping not only workflows but also team dynamics and organizational decision-making. Building on this literature, our study investigates the tensions, conflicts, and considerations that arise when UX practitioners adopt AI in collaborative settings. Following the multi-scalar format introduced in prior work at the intersection of UX and AI \cite{10.1145/3643834.3660720}, our study examines these dynamics at the individual, team, and organizational levels.

\subsection{Adoption-Centered Perspective in HCI}

Early HCI scholarship largely operated within a user-centered paradigm, emphasizing usability, efficiency, and satisfaction in interaction design and evaluation. Adoption, however, has long been theorized beyond immediate interaction. Rogers’ \textit{Diffusion of Innovations Theory} \cite{rogers2014diffusion} defines diffusion as the process through which an innovation is communicated over time among members of a social system. This account highlights that adoption is shaped not only by individual attitudes but also by social influence, organizational contexts, and broader institutional dynamics. It provided an early foundation for HCI to consider technology use as a process embedded in communities and organizations rather than as isolated user decisions. 

Hornbæk et al. \cite{10.1145/3127358} compared the Technology Acceptance Model (TAM) \cite{davis1989technology} with UX models \cite{hassenzahl2006user, hassenzahl2004interplay, o2010influence}, showing how these frameworks offer complementary but partial perspectives on adoption. TAM, rooted in management information systems, predicts adoption through constructs such as perceived usefulness and ease of use. UX models emphasize the experiential qualities of interactive products, such as whether a smartphone feels “classy” or “novel,” and their consequences for engagement. More recent work, such as the unified theory of acceptance and use of technology (UTAUT) model, expands these perspectives by incorporating organizational and contextual factors. It shows that adoption is shaped not only by individual attitudes, but also by facilitating conditions (e.g., infrastructure, training) and social influence \cite{dwivedi2019re}. Diffusion theory, TAM, UX, and UTAUT models together suggest that adoption is not reducible to usability alone but must be understood as a process shaped by prediction, perception, and social context.

In light of this, recent HCI research has explicitly articulated an adoption-centered perspective. Lindley et al. \cite{10.1145/3025453.3025742}, in Implications for Adoption, extend the tradition of “implications for design” by urging researchers to consider how technologies might be sustained and integrated after the prototype stage. They demonstrate how methods such as design fiction can help anticipate long-term consequences of adoption and foreground the situated relationships technologies foster. Similarly, Chilana et al. \cite{10.1145/2702613.2724724} present a case study tracing how the research prototype LemonAid evolved into the commercial product AnswerDash. Their analysis shows that while user-centered evaluation supported early success, the product’s viability ultimately depended on the needs of other stakeholders (e.g., administrators, buyers, and investors), who determined whether adoption would occur. These studies highlight the importance of understanding the broader ecosystems of adoption that shape whether innovations endure and create impact.

Extending this line of inquiry, scholars in HCI and adjacent fields have begun to examine responsible AI adoption \cite{10.1145/3706598.3713184} and how organizational contexts mediate both opportunities and challenges in integrating AI \cite{AIAdoptionAcrossMission-DrivenOrganizations}. For example, a recent think tank white paper \cite{AIAdoptionAcrossMission-DrivenOrganizations} highlighted that mission-driven organizations are increasingly exploring AI to enhance impact across operations and communication, yet continue to face barriers (e.g., leadership skepticism, ethical dilemmas, fragmented data, and dependence on large technology providers), factors that complicate strategic and values-aligned adoption \cite{10.1145/3706598.3713978}. Building on this trajectory, our study does not evaluate AI adoption in terms of predictive success, but rather investigates how adoption is actively deliberated within design practice and what tensions and considerations emerge in UX design practices.

%"That persepctive assets that what we would generally think of as cold computational processes––perceiving, thinking reasoning, decision making, categorizing––are shot through with values, needs, desires, and goals. We do not perceive an objective representation of the world; rather, we perceive a unique version colored by our unique desires and values as experienced in the situation we are engaged in. This means that in experience there is no given system activity" [tech as experience]

\section{Method}

\subsection{Study Procedures}
We aim to understand how UX designers experience and negotiate AI adoption in their everyday work practices. This study included two components: (1) a virtual group design workshop and (2) follow-up one-on-one interviews. To capture both collective deliberation and individual reflection, we adopted a mixed-method approach combining co-design workshops and follow-up interviews. Workshops allowed us to observe how designers collaboratively reasoned about AI adoption in the flow of their everyday work, revealing group dynamics, negotiations, and tensions. The follow-up individual interviews served as a complementary method to explore individual experiences and reflections that may not be openly expressed in group settings, including their perspectives on interpersonal dynamics, organizational pressures, and emotional responses.

\subsubsection{Recruitment}

\begin{table*}[t]
  \centering
  \caption{Participant Demographics and Group Composition}
  \label{tab:participants}
  \small
  \begin{tabularx}{\textwidth}{l|l|l|X|l|l|c}
    \toprule
    \textbf{Group} & \textbf{Sector} & \textbf{No} & \textbf{Job Title} & \textbf{Gender} & \textbf{Race} & \textbf{Years of Work Experience} \\
    \midrule
    \multirow{3}{*}{Group 1} & Finance & P1 & Senior Product Designer & Male & Asian & 6 \\
                             &        & P2 & Senior UX Designer & Female & Asian & 3 \\
                             &        & P3 & UX Designer & Female & Asian & 5 \\
    \midrule
    \multirow{3}{*}{Group 2} & Finance & P4 & Senior UX Designer & Male & Mixed & 4 \\
                             &        & P5 & UX Researcher & Male & North African & 5 \\
                             &        & P6 & Sr. Manager, User Research & Male & White & 10 \\
    \midrule
    \multirow{4}{*}{Group 3} & IT & P7 & Sr. User Experience and Design & Female & White & 25 \\
                             &    & P8 & Senior Product Designer & Male & Mixed & 10 \\
                             &    & P9 & UX design intern & Female & Asian & 1 \\
                             &    & P10 & Product designer & Female & White & 6 \\
    \midrule
    \multirow{3}{*}{Group 4} & Healthcare & P11 & Senior Experience Designer, Researcher and Strategist & Male & Asian & 9 \\
                             & Consulting & P12 & Senior Consultant & Female & Black & 5 \\
                             & Retail     & P13 & Sr. UXR Researcher & Female & Asian & 5 \\
    \midrule
    \multirow{2}{*}{Group 5} & IT & P14 & Product Designer \& Growth Manager & Male & Asian & 4 \\
                             & Healthcare & P15 & Product Designer & Male & Asian & 5 \\
    \bottomrule
  \end{tabularx}
\end{table*}

Between May and July 2025, we conducted a series of virtual design workshops with 15 participants (see Table~\ref{tab:participants}) from startups and mid- to large-sized technology companies based in the United States, Canada, and South Korea. We recruited self-identified designers in roles such as UX designers, design consultants, and UX/design researchers, working across sectors including finance, healthcare, education, consulting, and IT/AI. Participants were recruited through university networks, social media, word of mouth, and the research team’s professional connections, including snowball sampling.

We distributed a sign-up survey collecting information on industry sector, job title, gender, race, years of professional experience, and contact details. This information was used to organize participants into groups with shared professional or contextual backgrounds and to contact our participants.

Workshops were conducted in five groups. The first three groups consisted of participants from the same organization. The remaining two groups were intentionally assembled based on shared work contexts: the fourth group included product designers working at startups, while the fifth group consisted of UX design–related professionals with a common academic background.

To protect privacy, we have anonymized all personal identifiers, including names and organizational affiliations. Each participant received a 100 USD compensation in cash or as an Amazon gift card upon completing the study. One participant did not take part in the follow-up interview, and three participants voluntarily declined compensation. The study received approval from our university’s Institutional Review Board.

\subsubsection{Design Workshop}

We conducted virtual design workshops to explore how UX professionals reflect on, evaluate, and deliberate AI adoption in their practices. Each session lasted approximately 2.5 hours and was conducted via Zoom, with collaborative activities facilitated on Miro (see Figure \ref{fig:miro} in Appendix \ref{appendix:materials}). Sessions were recorded with participant consent.

Each workshop began with a framing discussion and reflective warm-up (up to 30 minutes), during which participants introduced themselves and shared their general attitudes toward AI, including any personal experiences with choosing not to use AI tools in their work. This initial conversation helped contextualize participants’ values and prompted reflection on AI use as a deliberate and situated practice.

Participants then moved through two core activities, each designed to scaffold deeper reflection and insight:

\paragraph{Activity 1: Scenario Building}
Each participant engaged in a scenario-building exercise to explore how organizational dynamics shape deliberations around AI adoption. This activity was designed to elicit reflection on values, constraints, and decision-making processes in workplace contexts. Participants selected from a set of three hypothetical organizational scenarios (see Appendix \ref{appendix:materials}) involving AI adoption (e.g., top-down AI mandates, compliance-sensitive client work, or team ambivalence). We designed these scenarios to reflect tensions commonly encountered in AI adoption and use in real-world organizational contexts, drawing from themes identified in prior research with UX professionals \cite{cha25chi}. Rather than centering on specific AI tools, the scenarios were crafted to surface underlying dynamics, such as managerial pressure, regulatory constraints, and conflicting team attitudes, that shape deliberation around AI use. This approach enabled participants to engage with plausible yet abstracted situations, prompting reflective discussion on values, constraints, and decision-making processes without requiring personal disclosure. Each participant was asked to choose one scenario.
This flexible assignment allowed participants to anchor their responses in contexts they felt best equipped to express, while still supporting a diversity of perspectives across the session. Participants were also encouraged to embed elements of their own organizational realities into the scenario-building process, allowing for grounded yet imaginative reflection on how such tensions might play out in their specific contexts. Although the scenarios were hypothetical, the imaginative outputs were treated as reflective of participants’ professional realities and values, as they often embedded elements from their own organizational contexts.

To guide participants through their scenario-building process, we provided a five-stage adoption framework—\textit{Discuss, Explore, Test, Select, and Adapt}—adapted from Rogers’ theory of adoption of innovations \cite{miller2015rogers}. Each stage was designed to scaffold participants’ thinking around the sociotechnical dynamics that shape AI adoption decisions in design practice. We translated Rogers’ abstract phases into more accessible, design-oriented terms (Details in Appendix \ref{appendix:materials}), each accompanied by reflection prompts to support structured yet flexible exploration.

For each stage, we provided the same guiding questions and associated prompts across the three scenarios which provided a clear structure for participants to articulate concrete actions, concerns, and evaluative criteria across the adoption process. At the same time, the flexible ``fill in'' nature (via sticky notes) of the activities welcomed participants to generate their own probing questions at each stage capturing emerging concerns or insights, where they often responded to their own questions as part of their scenario development. After working independently on each scenario, participants shared their situated scenarios and broader reflections  with the group and the other participants commented each other's stories. 

\paragraph{Activity 2: Surfacing and Structuring Critical Questions}
To continue the considerations on their own work experiences, we asked participants to zoom out from their scenario and think about key issues, challenges, and questions that emerged during their exploration of AI adoption. They began by working individually to generate questions and reflections about patterns they noticed in their day to day work both from their own experiences and from insights prompted by other's stories. After, they shared and discussed them with the entire group which allowed them to start collectively consider how and when such questions should be raised in real-world workflows. This structure aimed to explore the appropriate form, moment, and responsibility for asking these deliberative questions. For instance, participants considered whether questions about AI use should be embedded in design reviews and cross-functional planning meetings, and who should be responsible for initiating them. This part of the session helped situate the reflections, questions, and concerns in practical, organizational terms. 

\subsubsection{Follow-up Individual Interviews}

To further explore personal reflections and perspectives not fully captured in the group-setting workshop, we conducted one-on-one semi-structured interviews as a follow-up to the workshops. Each session lasted approximately 30 minutes and was conducted via Zoom. With participant consent, all interviews were recorded. The interview protocol covered six key areas (Details in Appendix \ref{appendix:interviewguide}): (1) participants’ background, including their organizational context, role, and working style; (2) a walk-through of their actual AI adoption process; (3) reflections on that process, including challenges and decision points; (4) feedback on the workshop experience; (5) broader reflections on AI in their professional practice; and (6) perspectives on the future of the UX role in relation to AI.

\subsection{Data Analysis}
We collected multiple forms of data during the workshops, including sticky notes from Miro boards, Zoom chat logs, and audio recordings. All workshop and interview recordings were transcribed verbatim. The lead author reviewed each transcript to verify the accuracy of the automated transcriptions and corrected any errors.

We employed reflexive thematic analysis \cite{braun2019thematic, braun2019reflecting}, which emphasizes the researcher’s active role in knowledge production. In this approach, themes are not simply discovered but are constructed through the interplay of data, the analytic process, and the researcher’s subjectivity.

Our analysis followed an inductive approach with an interpretivist stance \cite{Robert_Evaluating}, centering participants’ lived experiences and interpretations while acknowledging our own positionality as researchers and how it may shape the analysis. The analysis proceeded in multiple rounds. Two members of the research team reviewed the recordings and read each workshop and follow-up interview transcript multiple times, beginning with familiarization. One team member wrote analytical memos while watching the video recordings. Another team member conducted a conventional thematic analysis using Atlas.ti to code the transcripts, which also helped surface the main quotes presented in the Findings.

Based on the codes and memos, the full research team met regularly to discuss emerging insights, resolve ambiguities, and collaboratively define themes. We identified initial categories along with descriptions and illustrative examples. These included: (1) motivations and aims, (2) participants' professional contexts, (3) adoption processes, (4) considerations for adoption and usage, (5) reactions to and observations following AI adoption, (6) needs and pain points related to AI use, and (7) reflections on AI usage and adoption.

These categories were discussed and iteratively refined through a collaborative process of meeting, diverging, and synthesizing, resulting in the key insights presented in our Findings. Through this process, we came to recognize that AI adoption is deeply shaped by where designers are situated across multiple scales. This insight informed the overall structure of our findings.

\subsection{Positionality}
We are qualitative and design researchers working within an interpretivist paradigm \cite{Robert_Evaluating}. The lead author previously worked as a UX designer at an ML company, and this industry experience informed the study’s framing and methodological choices. Her familiarity with UX workflows and terminology helped build rapport with participants and supported nuanced interpretations of their reflections on AI adoption. The second author is a design researcher with experience working as a participatory designer within organizational settings, where she has used qualitative methods such as workshops, interviews, and observations to support the development of new products and services for both internal and external stakeholders. In her scholarly work, she draws on feminist theory to critically engage with her interlocutors’ everyday experiences of interacting with and relying on infrastructures. The last author combines expertise in science and technology studies, design, and HCI to study how designers attempt to attend to issues related to the social values that are promoted by technology and how they navigate social and organizational barriers in their workplaces. These experiences and perspectives informed the authors' analysis of the interviews. Our combined training and practice attuned our analysis to the relational, affective, and invisible forms of labor involved in adopting and resisting AI tools, and to how power, care, and obligation shape designers’ decisions about when and how to use AI. This shared background sensitized us to value conflicts in participants’ accounts, such as tensions between managerial metrics and designers’ own commitments to rigor and accountability. It also encouraged us to read participants’ narratives in relation to broader sociotechnical and organizational contexts rather than treating their choices as purely individual preferences.

\section{Findings: Negotiating the Values of AI Adoption}

In this section, we present our findings on the dynamics of AI adoption, showing how designers’ different positionalities emerged across individual, team, and organizational scales (summarized in Table \ref{tab:findings}). The central question was not simply whether AI adds value, but what ``value'' actually means, and how its meaning shifts depending on role, context, and priorities. As individuals, designers assessed AI in terms of efficiency, skill development, and long-term professional worth. At the team level, they grappled with how AI reshaped responsibility, rigor, and collaboration. At the organizational level, adoption was mediated by compliance requirements, formal structures, and industry pressures, where designers’ agency over tool use was entangled with culture and power dynamics. These positionalities surfaced in different moments yet overlapped in practice, and together they reveal that AI adoption is a site for an ongoing negotiation of competing values, stakeholder alignments, and evolving practices.

\begin{table*}[t]
\centering
\caption{Summary of Findings}
\label{tab:findings}

\begin{tabular}{p{0.95\linewidth}}
\toprule
\vspace{0.2em}
\textbf{Designers as End Users}

{\small
\begin{itemize}[leftmargin=1.2em]
    \item Participants adopted AI primarily to improve efficiency and productivity.
    \item Using AI encouraged reflection on long-term professional value and relevance.
    \item Efficiency was difficult to evaluate in practice despite being the main promise of adoption.
    \item Adoption involved substantial hidden labor including prompting, troubleshooting, verification, and correcting errors, which was often overlooked in formal assessments.
\end{itemize}
}
\\[0.4em]

\midrule
\vspace{0.2em}
\textbf{Designers Within Teams}

{\small
\begin{itemize}[leftmargin=1.2em]
    \item AI tools typically circulated through informal peer sharing rather than structured training, creating uneven adoption shaped by advocacy, role needs, and interpersonal dynamics.
    \item AI enabled some designers to perform tasks associated with other roles, raising concerns and emotional burdens about job stability for teammates with overlapping responsibilities.
    \item Participants saw AI as reshaping responsibilities and communication.
    \item AI shifted working norms and communication patterns, making transparency an important concern within teams.
    \item Expectations for upskilling varied across roles, generations, and work contexts.
\end{itemize}
}
\\[0.4em]

\midrule
\vspace{0.2em}
\textbf{Designers Within Organizations}

{\small
\begin{itemize}[leftmargin=1.2em]
    \item AI adoption was shaped by tension between leadership’s efficiency goals and institutional safeguards such as compliance, privacy requirements, and client expectations.
    \item Leadership emphasized productivity and cost savings, while approval processes moved slowly due to risk and compliance concerns.
    \item Adoption decisions frequently privileged business value over practitioners’ day-to-day needs.
    \item Organizational culture guided how adoption unfolded, reflecting existing norms, values, and routines.
    \item Even without decision-making authority, participants emphasized examining how adoption choices were made and what they signified for teams and work, including how AI’s value is defined, how long-term consequences are understood, and whose perspectives shape these judgments.
\end{itemize}
}
\\

\bottomrule
\end{tabular}

\end{table*}

\subsection{Designer as an End-User: Negotiating Efficiency and Self-Worth}

\subsubsection{Efficiency and Productivity as Primary Goals for Using AI}
Participants engaged with a diverse set of AI tools, ranging from stand-alone large language models (e.g., ChatGPT, LLaMA) and generative AI models (e.g., DALL-E) to AI features embedded within UX/design platforms (e.g., Maze, Dscout, Figma, Dovetail). These tools were employed for varied purposes, including transcribing interview data, supporting the development of interview guides, and assisting with UI design tasks. More broadly, participants applied AI across multiple stages of their work practices: generating interview questions, extracting user insights, creating 3D and graphic designs, and translating UI designs into front-end engineering implementations.

Participants consistently described efficiency and productivity as their primary motivations for adopting AI. They used tools to automate repetitive design work (e.g., resizing wireframes, producing high-fidelity variants), streamline documentation, and accelerate presentation-making. For some, AI felt truly transformative, with P8 noting that AI made them ``10 times better,'' highlighting its potential for significant productivity gains. P12 also emphasized that AI served less as a tool for production than as a mechanism for improving efficiency. In this sense, efficiency was framed not only as speed, but as a pathway to accomplishing more with limited time and resources.

\subsubsection{Tensions in Adopting AI Within the Self: Self-worth vs. Productivity.}
Yet, pursuing efficiency introduced new tensions. Adoption was rarely a clear yes/no choice; instead, participants described it as a spectrum of trade-offs. They emphasized that AI was meant to augment rather than replace their capabilities, but drew careful boundaries around what should remain human-led. P6 characterized AI as a ``digital prosthetic'' that extends their capabilities and helps them stay focused on their work, while others used AI for early drafts but excluded it from final deliverables. Some articulated beliefs about AI’s inevitability, with P13 warning, ``People who don’t use AI will get replaced by those who do.''

At the same time, participants reflected on risks to their long-term value. Even pro-AI participants, while emphasizing the importance of using AI properly, raised concerns about its potential to undermine their own necessity within the company. P14 admitted: 

\begin{quote}
    P14: ``Sometimes it makes me think, do I rely on AI tools too much? I’m afraid of losing my value within this organization… because I'm automating my workflow, using AI for transcribing, pulling insights from interviews, and writing PRD(Product requirements document)s for the dev team. It feels like I'm becoming too dependent on AI, and that it seems I might be easily replaceable by leadership. How can I protect myself from this perception? [...] I feel like I’m losing my core skills—so it’s kind of like I’m taking a gamble here.'' 
\end{quote}

This sense of dependency revealed deeper anxieties about being replaced by the very tools designed to enhance productivity. Similarly, junior designers worried that foundational ``muscles'' might atrophy, as P9 explained: ``Even though it’s faster and more efficient… we got to think about what muscles are not being used as a result of using these tools.'' Such concerns reflect an existential unease about the long-term value of their skills within organizations as automation accelerates. These accounts underscore that efficiency was not an unqualified good; it had to be weighed against values of skill-building, autonomy, and professional security.

\subsubsection{Evaluating Dynamics on AI's Efficiency}
While efficiency was the central promise for many individuals' use of AI, evaluating it in practice proved difficult. Participants struggled to measure AI’s impact beyond surface-level metrics. Many, including P15, preferred qualitative judgments, while P10 admitted relying on ``colleagues' informal, possibly inaccurate, perceptions'' because setting up formal tests was impractical. Many participants emphasized the need to evaluate AI’s productivity and impact, often suggesting KPIs (Key Performance Indicators) as a key method. They often measured tool uptake (e.g., sign-up rates, download numbers) rather than meaningful workflow change. As P7 noted, ``It’s just the presence or absence of experimentation at this point.''

%Participants emphasized the need to track improvements more effectively in order to assess AI’s true impact on their work. 
Participants drew attention to the hidden labor of adoption: repeated prompting, trial-and-error, and ongoing oversight to correct errors and hallucinations. P4 and P8 noted that while AI tools could be valuable, they demanded constant checking and adjustment, which increased rather than reduced workload. P5 stressed that without reliable accuracy, AI only added to manual effort. Similarly, P2 and P11 described the time lost to ``hidden trials'' before producing usable results. As P11 explained, ``ChatGPT is useful in many other ways, but I think I struggle with getting very relevant outcomes straight away. Like I have to do a lot of hidden trials more often.'' These forms of hidden labor \cite{10.1145/3613904.3642902} suggest that assessments of efficiency should account not only for visible outputs but also for the often-overlooked effort required to make AI tools workable.

Participants also discussed how AI adoption wasn’t limited to a single tool but involved a mix of tools used in different contexts. As P12 mentioned, ``We applied different tools in different contexts. It's very hard to evaluate a specific tool's impact on the overall process.'' These reflections underscore how participants approached AI not as a one-time adoption decision, but as an ongoing, evaluative process.

Generally, participants’ reflections reveal that efficiency was central but contested: sought as a value, yet entangled with concerns about accuracy, hidden labor, and professional development. Rather than a straightforward gain, efficiency had to be continually defined, tested, and balanced against other values.

\subsection{Designer Within a Team: Reconfiguring Work Practices, Responsibilities and Relationships}

As P12 observed, ``If it (a specific AI tool) hasn’t been circulating in the firm, it’s not going to be used.'' AI adoption was rarely an individual choice; it was embedded in team workflows and shaped by multiple stakeholders. While AI could accelerate individual tasks, its integration often reconfigured collaboration, redistributed work, and reshaped team dynamics. In this sense, adoption was less about automating work than about reorganizing how teams coordinated, how authority was distributed, and how relationships were maintained. These structural shifts frequently sparked tensions within teams.

% Adopting AI tools goes beyond automating tasks or increasing speed; it fundamentally reshapes how teams work, how responsibilities are divided, and how relationships are formed. The values and impacts that AI brings to workflows and work practices vary from person to person, depending on their role and perspective. Therefore, such changes frequently sparked tensions within teams.

% Participants reflected on the broader question of what constitutes ``value'' in adopting AI tools. While many saw AI as a way to expedite or support existing work, they questioned whether these tools provided sufficient benefits to justify transition costs and integration efforts. These reflections underscored that value was not absolute but negotiated across workflows, organizational contexts, and individual priorities. Practitioners often associated value with efficiency and productivity, yet stakeholders defined it differently, leading to divergent expectations and occasional tensions. Such differences produced emotional responses, resistance from team members (P8), or even stigmas associated with AI use (P11). As teams expanded and outputs scaled, these competing interpretations of AI’s relevance and impact risked becoming more pronounced (P9), underscoring the importance of treating adoption as context-dependent rather than universally beneficial.

Beyond these organizational dynamics, participants also grappled with what constitutes ``value'' in adopting AI tools. Some equated value with efficiency or faster outputs, while others weighed it against transition costs, integration burdens, or professional development priorities. Value, in this sense, was not absolute but negotiated across workflows, organizational contexts, and individual perspectives. Divergent definitions produced mismatched expectations, emotional responses, and even stigma around AI use (P8, P11). As teams scaled and outputs expanded, these competing interpretations of value risked becoming more pronounced (P9), underscoring that adoption should be understood as context-dependent rather than universally beneficial.

\subsubsection{Informal Diffusion}

As Rogers \cite{rogers2014diffusion} explains, diffusion refers to the process by which an innovation is communicated through specific channels over time within a social system. Similarly, in our study, AI tools and adoption decisions often spread informally rather than through structured training. Participants discovered tools through word of mouth, casual conversations, or ``lunch \& learns.'' As P13 noted, ``The use of AI tools that others have used that I could incorporate into my work in daily life. A lot of times, it’s word of mouth.''  

Early adopters or ``power users'' played a central role, demonstrating workflows, sharing results, and persuading colleagues through lived examples. P15 distinguished ``market approved'' tools such as ChatGPT from the ``false advertising'' of LinkedIn or YouTube promotions, relying instead on peer validation. Similarly, P14 described trying to persuade a skeptical manager by presenting case studies and demos, underscoring how persuasion required concrete outcomes rather than abstract claims.  

However, informal diffusion also generated tensions. Some viewed AI as a productivity booster, while others saw it as a shortcut undermining rigor. Managers like P12 reported having to ``push'' reluctant colleagues, particularly those who resisted on ethical grounds: ``I have some team members who are so averse to using AI because of ethical reasons… So, we literally have to push them to use it.'' By contrast, proactive adopters like P11, who experimented with tools independently, were celebrated as role models. These dynamics highlight how AI spread unevenly, shaped by advocacy, skepticism, and ethical debate.  

\subsubsection{Reshaping Role-specific Responsibilities}
Participants used AI to make their own design practices more efficient but also grappled with whether such adoption might threaten others’ job security. The introduction of LLMs blurred boundaries between UX and adjacent roles, including frontend engineers and content designers, heightening concerns about professional identity and stability. P2 and P3 voiced particular concern about the impact of AI on their counterparts’ roles. While acknowledging the benefits of AI in their own work, they recognized the anxiety felt by content designers as companies pushed for rapid adoption of LLMs. As P2 noted, ``There’s some sentiment around their job security... since LLM deals with human language, there’s a lot of potential to use it for content generation.'' P3 emphasized the emotional burden of such changes, particularly for those whose responsibilities overlapped most directly with AI.  

At the same time, they recognized content designers as indispensable in areas requiring judgment and legal compliance. As P3 explained, ``Content designers are responsible for making sure all the legal copy is included in designs. They play an important role in the workflow, partnering closely with the legal team.'' P2 similarly emphasized that content designers should be treated as integral partners rather than reduced to copywriters: ``At a human level, I would want to work with a human, not an AI.'' These accounts show how adoption reshaped professional boundaries while creating emotional burdens for those whose expertise overlapped with AI.  

Other participants envisioned more radical shifts. P14 argued that AI was expanding his responsibilities beyond design, enabling product managers and designers to take on tasks traditionally reserved for developers. Drawing on his own experience of submitting code to GitHub without formal programming expertise, he predicted that within a year junior developers could be displaced by product-focused professionals, and that the role of ``developer'' itself might eventually become obsolete.

\begin{quote}
P14: ``As a person who’s actively using a lot of AI tools, I would say less than a year, the junior engineer will be replaced easily. Just a few minutes ago, I submitted a PR to our product GitHub, even though I don’t know how to code. What I need is only a review from an engineer. As long as design, product vision, and coding can be aligned by one person, the workflow is much faster. If you think about previous workflows, sooner or later the word ‘developer’ will be gone.''
\end{quote}

While speculative, this perspective illustrates how participants perceived AI as a disruptive force capable of redefining responsibilities and reshaping the division of labor within teams. Observing such shifts, they expressed mixed emotional responses—ranging from excitement about new possibilities to anxiety over eroding roles and expertise. Participants stressed that value was not predetermined by role but emerged from how each individual assessed AI’s utility in relation to their own priorities, practices, and sense of contribution. These personal evaluative lenses meant that designers, researchers, and product managers often interpreted adoption differently, with one person’s judgment creating ripple effects across the team. For instance, adopting a tool to accelerate one task could implicitly signal that another role was redundant or less essential. P13 captured this challenge succinctly: ``What does this (AI adoption) mean to each person?'' For this reason, participants called for more structured reflection and prioritization practices, such as mapping trade-offs and systematically evaluating which tools to adopt and why (P1). Such processes, they argued, were essential to ensure adoption remained deliberate and aligned with team goals rather than driven solely by generalized promises of efficiency. This tension highlights the intersection between value in the economic sense of measurable productivity and values in the sense of broader social commitments such as collaboration, autonomy, and responsibility.

\subsubsection{Reshaping Workflows, Norms, and Communication}
Participants noted that AI not only altered workflows but also reshaped working norms and team communication. One central challenge, as P3 noted, was determining how to integrate AI effectively into existing processes. P12 emphasized that managers could not simply instruct designers to adopt AI; they had to experiment with tools themselves and provide clear, context-specific guidance. Similarly, P13 envisioned the ideal tool as ``a platform [AI tool] that fits into our everyday work so well that it runs without glitches.''

AI also introduced new responsibilities and emerging norms for collaboration, with transparency becoming a central concern. Some participants admitted that colleagues occasionally relied on AI outputs without verification, at times even boasting about fully AI-generated results. P4 and P6 cautioned that such practices could foster ``magic eight-ball thinking,'' or the tendency to accept AI-generated insights uncritically. P4 recalled instances where coworkers submitted strategy documents containing fabricated or non-functional links. Similarly, P6 described a case in which a UX researcher at a previous workplace fed large datasets into an AI tool without constraints, producing outputs that appeared plausible but were ultimately inaccurate. Drawing on his quantitative expertise, P6 identified these inconsistencies and underscored the need for oversight, awareness of AI’s limitations, and clear boundaries to mitigate hallucinations.

As P9 emphasized, transparency was essential for maintaining trust within teams:

\begin{quote}
P9: ``Another thing that I think is really important is once it gets implemented or continued to get used in the team is that it's important to communicate whether you used AI for a specific task before you kind of like pass it on to another member or if you're like working together on something, it's important to have that transparency of the fact that you used the AI tool for the specific tasks that you guys were both working on.''
\end{quote}

Participants worried that unverified or undisclosed AI use could undermine confidence in collaborative work. Several compared the solution to Wikipedia’s ``cite your sources'' ethos, calling for citation-like practices to distinguish human and AI contributions. The challenge was deciding when and how such disclosures should occur in order to preserve accountability and transparency with both colleagues and clients.

At the same time, many senior participants reflected on how efficiency gains could erode opportunities for discussion and negotiation. P2 noted a growing tendency to treat AI outputs as ``easy answers'' that bypassed productive conflict: 

 \begin{quote}
   P2: ``Other challenge was that there was this tendency of considering results from AI as an easy answer. Because it's like less political. So for instance, Like everyone has their own perspective. So we often have those conflicts. Or some or like most cases productive conflicts of like a designer things this way. However, the designer thinks this way, or this could be a conflict between stakeholders. However, I think there's a value of still like taking enough time Of doing that deep discussion to get to the results where everyone's happy where everyone's agreeing upon. Whereas with the adoption of AI. I found this tendency of like people just like whatever is created by AI like considering it as an answer or a shortcut For solutions like everyone's kind of like not eager to really debate or discuss or come up with their own ideas or Like dare to fight against in some way, but like people get lazy sometimes and like they just like consider AI answer to be like the best solution and not really like think further from there. So like, yeah, less passionate or less think thoughts included and design.''
 \end{quote}

P7 likened current practices to ``a bunch of kids with a bunch of candy,'' underscoring the need for reflection. Senior designers echoed this concern, warning that without structure, junior colleagues were especially vulnerable to over-reliance, completing tasks without critically evaluating outputs. Their limited experience made them prone to a ``convenience bubble,'' where quick answers were prioritized over collaboration, learning, and accuracy. These accounts highlight that AI adoption reshaped not only workflows but also the social fabric of teamwork, demanding new norms of oversight, disclosure, and debate to ensure that collaboration, critique, and learning were not eroded.

Perceptions of value differ across roles, experience levels, and organizational contexts, creating potential internal tensions. Although many practitioners prioritize productivity and efficiency, stakeholders define what is valuable differently, leading to variations in expectations. 

 \subsubsection{Upskilling Differs Across Roles and Levels}

Participants noted that AI adoption often carries an implicit expectation of ``upskilling,'' yet what counts as upskilling varied widely across contexts. P1 observed that working with older or less tech-savvy stakeholders could be challenging, as resistance to new technologies often slowed adoption. P9 emphasized that in multi-generational teams, individual experience and comfort with technology shaped both the pace and style of adoption:

\begin{quote}
P9: ``I think that would depend… We work in a very multi-generational world, and the sense of what upskilling means reflects differently across generations. The barrier for how easily someone will pick up and play around with a tool is not the same. Sometimes there’s genuine interest, and sometimes it’s just: ‘Hey everyone, we need to use our homegrown GenAI tool to move work faster.’ Everyone’s going to work at a different pace.''
\end{quote}

Since some team members readily experimented while others required structured guidance, participants emphasized that both interest and skill gaps had to be considered when integrating AI. The pace of adoption varied across projects and individuals, making ongoing assessment essential for effective and equitable use. P11 noted that voluntary exploration and testing were crucial, yet it was often difficult to determine when someone had genuinely upskilled, a process requiring sustained dedication and enthusiasm.

Upskilling related to AI was also described as bidirectional. Senior team members sometimes needed support to adopt AI tools, while senior designers bore responsibility for guiding junior colleagues in applying foundational research skills and maintaining critical thinking, ensuring they did not become overly reliant on AI. This reciprocal process helped teams strengthen capabilities without undermining core expertise.

Beyond generational differences, role-based priorities also shaped how participants evaluated AI’s value. Designers, researchers, and product managers brought distinct goals, and adoption in one area often rippled across the team. As P13 asked, ``What does this (AI adoption) mean to each person?'' A designer’s understanding of visual quality, for example, might differ from a product manager’s, meaning adoption was often judged through role-specific rubrics and priorities rather than shared criteria.

For these reasons, participants called for more reflection and structured approaches to adoption. P1 suggested prioritization mapping and trade-off analysis, emphasizing that ``there are a lot of different tools and different ways to use them,'' underscoring the need to carefully evaluate which AI tools to adopt and why. In this sense, participants framed adoption not simply as a quest for technical efficiency, but as a negotiation between efficiency as economic value and values such as collaboration, professional growth, and autonomy.

\subsection{Designer Within an Organization: AI Adoption as a Mirror of Culture}

Although AI tools are often introduced with the promise of efficiency and speed, their adoption is rarely straightforward. Organizational infrastructures impose constraints, and practitioners who will ultimately use the tools often have little authority over which ones are adopted. In most cases, decisions were driven by management, with designers providing feedback only within pre-defined boundaries. As a result, participants found themselves navigating organizational politics around AI adoption, even without formal decision-making power.  

Several emphasized that, while they could not directly influence whether a tool was adopted, they valued reflecting on the adoption process itself and its implications for their teams, workflows, and organizational culture. In this sense, deliberation was less about choosing tools than about shaping how adoption was positioned, evaluated, and integrated. Such discussions fostered a sense of alignment among stakeholders while mirroring broader cultural norms of organizational decision-making.

\subsubsection{Adoption Constrained by Organizational Infrastructure}

At the organizational level, AI adoption was shaped by a tension between efficiency goals and institutional safeguards. While leadership often framed AI in terms of productivity gains and cost savings, formal approval was slowed by compliance requirements, privacy concerns, and client sensitivities. Participants described leaders weighing potential benefits against regulatory constraints, budget considerations, and client expectations.

The process varied across company size and industry. Startups tended to embrace AI more openly, prioritizing speed and experimentation, whereas larger firms relied on compliance-heavy procedures requiring leadership approval, budget allocation, security reviews, and client alignment. In large firms, AI often existed in a ``gray zone'', used unofficially until it became too valuable to ignore. Participants from such firms, including P7 and P13, described approval processes that frequently outlasted the project timeline itself: ``Sometimes we get compliance approval in week 7 of an 8-week project'' (P13). These cumbersome timelines discouraged experimentation altogether.

Industry-specific constraints further shaped adoption. P14, working in IT, described relative freedom to experiment, while P15, working in healthcare, faced HIPAA\footnote{HIPAA is a U.S. health data privacy law.} compliance that significantly slowed adoption. Participants in client-facing industries such as finance or consulting reported additional limits when clients imposed outright bans. Vendor supply chains and external providers often dictated standards as well, illustrating how adoption was influenced as much by external forces as by internal priorities. At the same time, broader market and economic pressures continued to push organizations toward AI adoption, even when internal trust in the technology remained limited.

\subsubsection{Decision-Making Power}
Although teams often experimented with AI tools informally, with P11 recalling that he once ``tested a tool over the weekend,'' formal adoption was largely management-driven. Official approval required leadership sign-off, compliance clearance, or client alignment, leaving practitioners positioned as ``end-users'' rather than true decision-makers. As P4 and P5 reflected, adoption was ``something managers like P6 decided,'' underscoring their lack of agency.  

Participants, including P14, observed that such top-down dynamics reflected leadership’s focus on efficiency and business value, often at the expense of practitioners’ needs. Companies invested in tools even when productivity gains were unclear, raising questions about whether adoption was driven by optimism, external pressure, or strategic signaling. P7 described leaders pursuing innovation ``for innovation’s sake,'' while P10 wondered if adoption was ``just checking a box […] or actually helping us in a real measurable way.'' Similarly, P8 questioned whether business priorities outweighed user needs: ``The business just wants me to use AI to generate UI, and it’s currently trash. Should I just keep hammering it until some magic happens?'' These accounts suggest that adoption decisions were less about enabling practitioners to choose tools suited to their workflows, and more about decision-makers’ willingness to invest, sometimes with blind trust, in tools they believed would yield efficiency gains.

\subsubsection{AI Adoption as a Mirror of Organizational Culture}

Actual adoption trajectories often diverged from practitioners’ situated needs. Several participants reported feeling that their engagement with AI was driven less by necessity than by organizational expectations to adopt or showcase the technology. This dynamic revealed a broader divide over how ``value'' was defined. Leadership frequently wanted to evaluate value with efficiency gains and productivity metrics, while some workers viewed AI as contributing to deskilling, diminishing the craft of their work, or even threatening their professional roles. As a result, definitions of ``success'' were uneven, interpreted differently across hierarchical positions and organizational contexts, and became a source of ongoing tension.

Participants emphasized that AI adoption was not only influenced by organizational culture but also mirrored it. The ways in which adoption unfolded reflected existing norms, values, and practices. Adoption was not limited to task-level changes but also reshaped interpersonal relationships, making reflection and deliberation essential.

Several participants raised questions about scenarios in which their UX teams might lose the ability to individually opt in or out of AI use for specific tasks, reflecting autonomy, and the boundaries of organizational authority. P9, for instance, questioned how differing opinions about AI integration would be managed:

\begin{quote}
P9: “I was kind of thinking about how it would affect company culture, how workers actually work together and […] how they’d get along. […] Will team members have differing opinions about AI being integrated with the current workflows? […] What systems or practices will be put in place to resolve these conflicts? […] Workshops or discussions might help get these opinions out there and let teams assess whether using AI would be good for their workflow. […] I’m interested in thinking about how the company would combat these issues, because integration […] can be divisive and even create tension in the workplace.”
\end{quote}

Even without decision-making power, participants stressed the importance of reflecting critically on how adoption decisions were made and what these choices meant for their teams and work. Such reflection raised broader questions: What kinds of value does AI generate, and how should that value be measured? What are the long-term implications of embedding AI into professional practice? And whose perspectives ultimately shape these definitions?

These concerns circled back to questions of agency: could workers decline to use AI without facing peer pressure or coercion? P15 highlighted the role of leadership in shaping this environment, noting that leaders could either impose AI use or instead foster a safe space for dialogue by clearly explaining benefits and engaging teams in decision-making. Similarly, P7 emphasized the importance of measurement, asking not only ``how are we going to do this?'' but also how such assessments could be conducted in ways that gave workers control over what was being measured and how results were reported.

Other participants described a sense of being overwhelmed by the scale and transformative potential of AI. P12 contrasted AI with earlier transitions to cloud computing, framing AI as far more ``shape-shifting'' and destabilizing. They reflected on the challenge of finding their own place amid uncertainty:

\begin{quote}
P12: ``The emotion I have is I need to wrangle this thing in […] to figure out how to grow as I’m using it, but not be overcome by the fact that it has the power to do multiple different things. […] We all have a sense this is going to wipe out so many roles, but how do we find ourselves and shape shifts in the mix of all of it, so we can still reach the goal […] ?''
\end{quote}

Across these reflections, participants positioned adoption not as straightforward tool uptake but as a contested process shaped by competing values, organizational logics, and persistent uncertainties about necessity and worth. Rather than a neutral mechanism of efficiency, AI adoption became a cultural and relational practice—one that surfaced tensions over autonomy, professional identity, and the definition of value itself.

% Participants also raised a more fundamental concern: was AI being adopted because it genuinely addressed pressing needs, or because there was a perceived obligation to do so? For example, P15 described “going one level deeper” by asking what the hard facts of adopting a new tool were and what specific goals it was meant to achieve. Similarly, P10 reflected on the temporal trade-offs of adoption, questioning how long it would take for a tool to become genuinely usable and at what point the setup effort might outweigh its benefits:

% P10: “Have we now spent so much time setting up this tool and getting it to be right that it's just not worth it to use it anymore?”

% Such reflections highlight that AI adoption was not experienced as straightforward tool uptake, but as a contested process shaped by competing values, organizational logics, and persistent uncertainties about necessity and worth.

\section{Discussion}
Our findings highlight that what ultimately matters in the adoption and use of AI tools is not simply their technical functionality, but the roles they play within teams and organizations and the meanings people attach to them. Participants showed how discourses of efficiency often carried social dimensions (e.g., responsibility, trust, autonomy), revealing adoption as a process of reconfiguring who does what, who decides, and whose contributions count. Adoption also raised new questions about agency in shaping these decisions.

Participants further questioned the very motivations for adopting AI. Was AI being integrated because it genuinely addressed pressing needs, or because organizations felt a broader obligation to embrace technological promises? As P15 put it, adoption often meant ``going one level deeper'' to ask what concrete goals a new tool was meant to achieve. P10 similarly reflected on the temporal trade-offs of adoption, wondering whether ``we’ve now spent so much time setting up this tool and getting it to be right that it’s just not worth it to use it anymore.'' These reflections underscore that AI adoption was not experienced as straightforward tool uptake, but as a contested negotiation over value and values, raising questions about necessity, legitimacy, and the future role of human expertise. In what follows, we turn to the discussion to examine how efficiency is being reinterpreted, how AI reshapes practices and values in situated ways, and how adoption ultimately hinges on who decides when and how such tools are taken up.

\subsection{Understanding Efficiency through the Values of Value}
When AI is adopted in organizations, 
% \st{it is not simply layered onto existing practices. It actively reshapes labor, workflows, and how value is recognized.}
it is not simply added onto existing practices. As our findings show, it actively reorganizes labor, reshapes workflows, and alters which and whose values come to be valued. % how value is recognized, including which values come to be valued and whose.}
At the individual level, participants described AI in terms of productivity and skill-building, but these aspirations collided with values of autonomy, mastery, and long-term professional worth, raising questions about 
% \st{what counts as meaningful work}
what counts as meaningful work and who gets to define it. At the team level, AI adoption promised faster workflows and lighter workloads, yet these shifts unsettled commitments to rigor, shared responsibility, and transparent collaboration. At the organizational level, AI adoption was shaped by a tension between efficiency goals and institutional safeguards. Leadership often framed AI in terms of productivity gains and cost savings, yet formal approval slowed under compliance requirements, privacy concerns, and client sensitivities. This produced a definition of value that emphasized risk management and efficiency while sidelining professional identity and collaborative practices.

Across these scales (individual, team, and organization), participants showed that AI reconfigures not only how tasks are done but also how contributions are evaluated. Efficiency was no longer defined as just speed, but was also conceptualized to include responsibility in distributing 
% \st{workload}
labor, or trustworthiness in outputs. Social values such as rigor, autonomy, or professional security were reframed in the language of value—“what is AI worth to me, my team, or my company?” In this sense, adoption was less about whether AI creates value in a narrow economic sense, and more about how it reorganizes the conditions under which value itself is defined, measured, and legitimized.

Graeber’s anthropological account of value helps unpack this tension. In \textit{Toward an Anthropological Theory of Value} \cite{Graeber_2001}, he distinguishes between value in the economic sense (e.g., exchange, productivity, efficiency) and values in the social sense (e.g., ideals of what is good or desirable). Our data showed how these two registers were constantly \textit{entangled} rather than separable. This echos \citet{d2023data}'s arguments that metrics such as ``efficiency’’ are never purely economic but are already enmeshed with social and political values, and that recognizing this entanglement is key to understanding how power operates through data systems. Even when participants invoked economic language like efficiency or productivity, these terms carried conceptions of social values: efficiency was rarely just about speed or cost reduction, but also about transparency, trustworthiness, or autonomy. Conversely, social values such as rigor or professional identity were often articulated in the language of economic value, as questions of worth at individual, team, or organizational scales. \textbf{This interplay shows that efficiency is not a neutral descriptor but a contested site where economic value claims collide with social values.}

% \st{To make sense of how these values shift across contexts, we suggest future research could draw on the concept of handoff}
% \st{. Handoff refers to the role transitions and transfers of responsibility that occur as tasks or functions move between different actors.}
Understanding how these values shift across contexts is an important direction for future HCI research. As Mulligan et al. \cite{Mulligan_Nissenbaum_2020} emphasize, %such transfers 
the introduction of new technologies into existing systems 
are never simple operational substitutions; they are moments in which accountability, agency, and normative expectations are redistributed. In this vein, rather than presuming modular divisions of labor, we should question the assumption that swapping system components (in this case AI tools to augment labor) leaves ethical or political dimensions unchanged. Future work could examine how responsibilities become obscured, displaced, or reassigned, and surface the hidden labor and ethical stakes that emerge as AI systems are introduced into workplaces.

Extending this focus on redistribution, feminist and care-oriented perspectives invite questions about how relational accountability is reconfigured, rather than simply tracked, when work circulates between people and AI systems.
As Wieczorek et al. \cite{wieczorek2025architecting} argue, 
% \st{AI integration causes a reconfigurations that are not only structural but shift accountability, labor, and trust. When seen through this lens, AI adoption is not only about adding efficiency but about reconfiguring who does what, who decides, and whose contributions are recognized. Handoff draws attention to how AI systems redistribute responsibility among team members, reshape organizational politics, and reconfigure the social impacts of work.}
AI integration produces reconfigurations that are not only technical or structural but also shift accountability, labor, and trust; this resonates with ICTD work showing how care and coordination are redistributed within fragmented infrastructures \cite{karusala2023unsettling}.
Seen through this lens, AI adoption is not simply about adding efficiency but about reorganizing who does what, who decides, and whose contributions are rendered legitimate. 
This framing foregrounds how AI systems redistribute responsibility across teams, shaping who absorbs additional coordination work and whose labor becomes newly peripheral or newly visible. As we saw our when interlocuters shared concern for the security of their colleagues' jobs, AI-adoption is not simply about the asks to be completed but the relational and care work involved in daily work experiences.
%%
 
% \st{Grounded in past work in social studies of technology and values in design, this perspective disrupts the assumption that replacing one ``modular" component of a system with another leaves ethical and political dimensions untouched. Instead, it shows how the movement of functions across actors, whether between humans and AI tools, or between practitioners and management, redefines workflows, shifts power, and makes certain values visible, contested, or marginalized. Using handoff as a lens for AI adoption highlights how efficiency claims are intertwined with social values. This perspective draws attention to the ways teams negotiate questions of agency, responsibility, and value through the tools they take up. As our study demonstrates, adopting AI reshapes not only individual task management but also how collaboration unfolds, how responsibilities are distributed, and how decisions are made. Attending to handoff practices thus helps us see adoption less as a matter of technical efficiency and more as an ongoing reconfiguration of work and relationships.}

% \st{Considering how a handoff might play out in practice opens up new opportunities to consider how AI could be leveraged in teams.}
Rather than focus on efficiency and improvement of particular tasks, it invites us to reflect on how AI also reshapes the affective and experiential dimensions of teamwork. In our findings, we see how AI adoption creates uncertainty where workers unsure of their place within the team, their future career trajectory, and even their sense of professional identity. This ambiguity generates emotional tolls (e.g., anxiety, loss of trust, and even competition among teammates) as workers attempt to renegotiate their roles, relevance, and relationships. In creative domains especially, where collaboration and trust are key, the blurring of workflows and values can create environments of latent tension rather than 
% \st{collective confidence} 
shared assurance.

% \st{At the same time, these frictions open possibilities for reorienting AI toward relational practices. If AI is not only understood as an individual extension but as a relational actor, it could be designed to support interdependence within teams rather than exacerbate fragmentation. This opens opportunities to imagine AI as a mediator of values, a facilitator of shared workflows, or even a collaborator attentive to the dynamics of trust and care among workers. In this sense, the future of AI in collaborative settings may hinge not on its capacity to replace, but on its ability to sustain and enrich relationality.}

These frictions suggest the need to examine AI not solely as a tool, but as a sociotechnical force that reorganizes how people relate to one another and how their work is valued. AI integration influences the less visible infrastructures of workplace life like trust, responsibility, professional worth, and the distribution of recognition. As roles shift and tasks are delegated by management, workers may feel their expertise is devalued or that accountability becomes (especially as they navigate working individually vs. across teams), raising questions about who is responsible when decisions go wrong or when creative ownership is contested. For these reasons, designing AI systems for collaborative environments requires attention to whether they reinforce fragmentation, diminish workers’ contributions, or unsettle professional identity, versus whether they sustain interdependence, clarify responsibility, and support a sense of shared purpose. Understanding these broader relational and organizational consequences is critical for imagining forms of AI adoption that support—not destabilize—the social and professional conditions that make collective work possible.

%These frictions point to the need to examine AI not solely as a tool, but as a sociotechnical force that reorganizes how people relate to one another and how their work is valued. AI integration influences not only coordination and communication but also the less visible infrastructures of workplace life, including trust, responsibility, professional worth, and the distribution of recognition. As roles shift and tasks are delegated, workers may feel that their expertise is devalued or that accountability has become unclear, which raises questions about who is responsible when decisions go wrong or when creative ownership is contested. Designing AI systems for collaborative environments therefore requires attention to whether they reinforce fragmentation, diminish workers’ contributions, or unsettle professional identity, versus whether they sustain interdependence, clarify responsibility, and support a sense of shared purpose. Understanding these broader relational and organizational consequences is critical for imagining forms of AI adoption that support, rather than destabilize, the social and professional conditions that make collective work possible.

\subsection{Values of Adoption from Multiple Perspectives}

Our findings show that AI adoption is frequently promoted under the promise of productivity and efficiency. Yet in practice, organizations rarely assess whether such efficiency is ever realized, echoing longstanding concerns in organizational theory about problems of measurement and identification. As Syverson notes in his survey of productivity research, ``no potential driving factor of productivity has seen a higher ratio of speculation to empirical study'' \cite{syverson2009determines}. Also, in pervious organizational studies, rigorous empirical assessments remain scarce, and large-scale experiments within firms are especially rare \cite{learned2009impact}. Our study echoes these concerns: participants described AI adoption as similarly promoted under the banner of efficiency and corporation's speculative initiatives, however, rarely subjected to meaningful evaluation, often generating tensions and conflicts rather than clear productivity gains.

This lack of meaningful evaluation is also related to how adoption unfolded on the ground. Rather than being determined by practitioners themselves, AI use was largely conditioned by organizational infrastructures and leadership priorities. While management tended to frame “success” in terms of efficiency and productivity, workers more often experienced adoption as deskilling, a loss of craft, or even a threat to their professional roles. This uneven framing created ongoing tensions, with adoption processes both shaped by and reinforcing existing organizational cultures and norms.

Participants recognized AI adoption as a site of tension: calls for efficiency aligned closely with managerial perspectives rather than those of practitioners. This dynamic resonates with long-standing critiques in management science, where ``efficiency'' has historically been linked to managerial control, narrowing worker discretion, privileging quantifiable activities, and subdividing labor \cite{greenbaum1979name}. As Graeber observed, ``‘Efficiency’ has come to mean vesting more and more power to managers, supervisors, and presumed ‘efficiency experts,’ so that actual producers have almost zero autonomy'' \cite{Graeber_2019}. In this light, efficiency often functioned less as an indicator of genuine productivity gains than as a justification for heightened managerial authority. Historical critiques of technology at work sharpen this interpretation: technological innovation has rarely been introduced independently of the people implicated in its use \cite{Barry_1981}, while Greenbaum noted that workplace rules and procedures typically prioritized control and standardization over efficiency or creativity. She further argued that simplified programming languages allowed management to reduce reliance on skilled workers, increasing workloads without improving job quality—“rather than more ‘creative’ work there is now just more work for each person to do” \cite{greenbaum1979name}. Our study extends these critiques into the present, showing how efficiency-driven framings of AI adoption similarly obscure the ways in which AI increases burdens on practitioners while limiting their autonomy, thereby reinforcing organizational control rather than genuinely enhancing creativity or productivity.

Building on these perspectives, our findings suggest that future research should investigate not only mechanisms that enable workers to exercise meaningful decision-making power in the adoption of AI, but also the structural limits of such efforts. Worker agency in design work is often tightly circumscribed by broader political economies of automation, outsourcing, precarity, and managerial control, as well as by legal and organizational arrangements that privilege employers’ interests, echoing D’Ignazio \& Klein’s critique of how power structures limit participation and possibility \cite{d2023data}.
% \st{HCI researchers, in particular, are well positioned to contribute by drawing on long-standing traditions such as }
Within these constraints,
participatory design (PD) traditions such as
Scandinavian PD (e.g., \cite{floyd1989out, gregory2003scandinavian, schuler1993participatory}) 
% \st{which is politically motivated to promote workplace democracy and support worker well-being. Such approaches help workplaces, organizations, and communities build the collective resources needed to decide if}
offer one set of resources for experimenting with more democratic forms of workplace governance by asking when, and how AI should be integrated into specific contexts.
However, such approaches are politically situated practices that might be enrolled in consultation exercises or ``participation washing'' if they are not linked to broader forms of worker organization and collective bargaining. The stakes are not just about who gets to participate on what terms, but also whose values are promoted and reflected in workplace institutions and systems of power.
For instance, the recent negotiations by the Writers Guild of America (WGA) \cite{WGAAI} demonstrate how creative workers, backed by union power, organized to set boundaries on the use of AI in scriptwriting to ensure that AI could not replace human authorship but might be used under conditions they define. 
% \st{Such cases highlight the potential for worker-led governance to serve as models for shaping AI adoption in ways that protect professional identity and redistribute decision-making power. Pursuing such inquiry not only moves beyond efficiency-driven framings but also contributes to broader efforts in HCI to redistribute agency and voice in the shaping of technological futures, aligning with recent ideas of AI counter-governance}
We therefore understand worker-led AI governance not as a ready-made solution but as a contested arena in which the terms of AI adoption are fought over, aligning with recent calls for AI counter-governance that foreground oppositional, resistant, and solidaristic practices, rather than assuming that better design alone can resolve the tensions produced by AI integration \cite{Sun_2023}.

\section{Conclusion}

Our study demonstrates that understanding AI adoption requires moving beyond technical efficiency to grapple with the deeper sociological and ethical dimensions of value, contributing to a broader turn in HCI research focusing on issues related to adoption rather than use. As Graeber \cite{Graeber_2001} observes, attempts to theorize value in isolation by only focusing on economic sense often falter when neglecting its other dimensions. Our work similarly demonstrates that AI adoption cannot be understood solely in terms of technical efficiency (corroborating HCI and adjacent research that critically analyzes the politics of AI). Efficiency invokes economic metrics of productivity; concerns about autonomy and rigor reflect sociological values; and the very language used to frame adoption signals linguistic distinctions that shape meaning. Therefore, these entanglements reveal efficiency as a contested site where multiple notions of value collide.

Our findings suggest that what ultimately matters in the adoption and use of AI tools is less their technical functionality than the roles they assume within teams and organizations, and the meanings people attach to them. While participants frequently described adoption in terms of efficiency, such framings were rarely grounded in systematic evaluation. Instead, AI reshaped workflows, reconfigured relationships with colleagues, and introduced new questions of agency in decisions about when and how tools should be taken up. Discourses of efficiency thus carried social dimensions (e.g., responsibility, trust, and autonomy) revealing adoption as a process of renegotiating who does what, who decides, and whose contributions count.

Looking ahead, we suggest that the critical issue is not whether AI enhances productivity, but how it redistributes responsibility and reshapes organizational politics, and alters how value is recognized. 
% \st{Through the lens of handoff \cite{Mulligan_Nissenbaum_2020}, we see AI as simultaneously producing uncertainty and identity struggles for workers while also opening possibilities for it to act as a relational mediator that fosters interdependence and care.} 
Future research should attend to how these shifts produce uncertainty, unsettle professional identity, and redistribute accountability, often in ways obscured by the name of efficiency.
At the same time, efficiency may continue to serve as a rationale for managerial control, often disregarding workers’ own definitions of productivity. Future research should therefore develop mechanisms that strengthen worker agency in deciding if, when, and how AI is integrated ensuring that adoption is guided by social values and organizational cultures.

\begin{acks} 
Thank you to the participants for sharing their experiences with us. We would like to thank the anonymous reviewers, and the Georgia Tech Creating Ethics Infrastructures Lab members for their insightful discussions in shaping this work. Thanks to Pooja Casula, Sarah Mathew, and Sylvia Janicki for their constructive feedback and comments. This work was partly supported by a seed grant from the Atlanta Interdisciplinary AI Network.
\end{acks}
\bibliographystyle{ACM-Reference-Format}
\bibliography{Reference}

@String{Computing = "Computing" }

@String{Computer = "{IEEE} Computer" }

@String{Macmillan = "Macmillan" }

@String{Springer = "Springer-Verlag" }

@techreport{AIAdoptionAcrossMission-DrivenOrganizations,
  author       = {Dalia Ali and Muneeb Ahmed and Arfa Khan and Hailan Wang and Sunnie S. Y. Kim and Orestis Papakyriakopoulos},
  title        = {AI Adoption Across Mission-Driven Organizations},
  institution  = {TUM Think Tank},
  type         = {White Paper},
  year         = {2025},
  url          = {https://tumthinktank.de/en/output/ai-adoption-across-mission-driven-organizations/},
  month        = aug,
  day          = {5},
  note         = {Available under CC BY-ND 4.0 license},
}

@inproceedings{10.1145/3706598.3713184,
author = {Bhatnagar, Tigmanshu and Omar, Maarya and Orlic, Davor and Smith, James and Holloway, Catherine and Kett, Maria},
title = {Bridging AI and Humanitarianism: An HCI-Informed Framework for Responsible AI Adoption},
year = {2025},
isbn = {9798400713941},
publisher = {Association for Computing Machinery},
address = {New York, NY, USA},
url = {https://doi.org/10.1145/3706598.3713184},
doi = {10.1145/3706598.3713184},
booktitle = {Proceedings of the 2025 CHI Conference on Human Factors in Computing Systems},
articleno = {1098},
numpages = {17},
location = {
},
series = {CHI '25}
}

@incollection{braun2019thematic,
  title={Thematic analysis},
  author={Braun, Virginia and Clarke, Victoria and Hayfield, Nikki and Terry, Gareth},
  booktitle={Handbook of research methods in health social sciences},
  pages={843--860},
  year={2019},
  publisher={Springer}
}

@article{braun2019reflecting,
  title={Reflecting on reflexive thematic analysis},
  author={Braun, Virginia and Clarke, Victoria},
  journal={Qualitative research in sport, exercise and health},
  volume={11},
  number={4},
  pages={589--597},
  year={2019},
  publisher={Taylor \& Francis}
}

@article{Robert_Evaluating,
author = {Soden, Robert and Toombs, Austin and Thomas, Michaelanne},
title = {Evaluating Interpretive Research in HCI},
year = {2024},
issue_date = {January - February 2024},
publisher = {Association for Computing Machinery},
address = {New York, NY, USA},
volume = {31},
number = {1},
issn = {1072-5520},
url = {https://doi.org/10.1145/3633200},
doi = {10.1145/3633200},
journal = {Interactions},
month = jan,
pages = {38–42},
numpages = {5}
}

@incollection{goransdotter2022designing,
  title={Designing Together: On Histories of Scandinavian User-Centred Design},
  author={G{\"o}ransdotter, Maria},
  booktitle={Nordic Design Cultures in Transformation, 1960--1980},
  pages={157--177},
  year={2022},
  publisher={Routledge}
}

@article{muller1993participatory,
  title={Participatory design},
  author={Muller, Michael J and Kuhn, Sarah},
  journal={Communications of the ACM},
  volume={36},
  number={6},
  pages={24--28},
  year={1993},
  publisher={ACM New York, NY, USA},
  doi={https://doi.org/10.1145/153571.255960}
}

@book{schuler1993participatory,
  title={Participatory design: Principles and practices},
  author={Schuler, Douglas and Namioka, Aki},
  year={1993},
  publisher={CRC press}
}

@inproceedings{10.1145/3411764.3445237,
author = {Lin, Cindy and Margot Lindtner, Silvia},
title = {Techniques of Use: Confronting Value Systems of Productivity, Progress, and Usefulness in Computing and Design},
year = {2021},
isbn = {9781450380966},
publisher = {Association for Computing Machinery},
address = {New York, NY, USA},
url = {https://doi.org/10.1145/3411764.3445237},
doi = {10.1145/3411764.3445237},
booktitle = {Proceedings of the 2021 CHI Conference on Human Factors in Computing Systems},
articleno = {595},
numpages = {16},
location = {Yokohama, Japan},
series = {CHI '21}
}

@article{smith2015continuity,
  title={Continuity and change in labor process analysis forty years after labor and monopoly capital},
  author={Smith, Chris},
  journal={Labor Studies Journal},
  volume={40},
  number={3},
  pages={222--242},
  year={2015},
  publisher={SAGE Publications Sage CA: Los Angeles, CA}
}

@misc{Sun_2023, title={AI Countergovernance}, url={https://www.midnightsunmag.ca/ai-countergovernance/}, abstractNote={Blair Attard-Frost on resistance to artificial intelligence, backlashes against the governance of AI systems, and possibilities for bottom-up AI governance led by communities and workers.}, journal={Midnight Sun}, author={Sun, Midnight}, year={2023}, month=dec, language={en-US} }

@article{10.1145/2832117,
author = {Ekbia, Hamid and Nardi, Bonnie},
title = {The political economy of computing: the elephant in the HCI room},
year = {2015},
issue_date = {November - December 2015},
publisher = {Association for Computing Machinery},
address = {New York, NY, USA},
volume = {22},
number = {6},
issn = {1072-5520},
url = {https://doi.org/10.1145/2832117},
doi = {10.1145/2832117},
journal = {Interactions},
month = oct,
pages = {46–49},
numpages = {4}
}

@inbook{Mulligan_Nissenbaum_2020, edition={1}, title={The Concept of Handoff as a Model for Ethical Analysis and Design}, ISBN={9780190067397}, url={https://academic.oup.com/edited-volume/34287/chapter/290662274}, DOI={10.1093/oxfordhb/9780190067397.013.15}, abstractNote={Abstract              This chapter introduces the concept of handoff, which offers a lens through which to evaluate sociotechnical systems in ethical and political terms. It is particularly tuned to transformations in which system components of one type replace components of another. Of great contemporary interest are handoff instances in which AI take over tasks previously performed by humans, for example, labelling images, processing and producing natural language, controlling other machines, predicting human action (and other events), and make decisions. Grounded in past work in social studies of technology and values in design, the handoff analytical model disrupts the idea that if components of a system are modular in functional terms, replacing one with another will leave ethical and political dimensions intact. Instead, the handoff lens highlights different ways that different types of system components operate and interoperate and shows these differences to be relevant to the configuration of values in respective systems. The handoff lens offers a means to make ethically relevant changes salient that might otherwise be overlooked.}, booktitle={The Oxford Handbook of Ethics of AI}, publisher={Oxford University Press}, author={Mulligan, Deirdre K. and Nissenbaum, Helen}, editor={Dubber, Markus D. and Pasquale, Frank and Das, Sunit}, year={2020}, month=jul, pages={232–251}, language={en} }

@article{gregory2003scandinavian,
  title={Scandinavian approaches to participatory design},
  author={Gregory, Judith},
  journal={International Journal of Engineering Education},
  volume={19},
  number={1},
  pages={62--74},
  year={2003},
  publisher={TEMPUS publications}
}

@article{floyd1989out,
  title={Out of Scandinavia: Alternative approaches to software design and system development},
  author={Floyd, Christine and Mehl, Wolf-Michael and Resin, Fanny-Michaela and Schmidt, Gerhard and Wolf, Gregor},
  journal={Human--computer interaction},
  volume={4},
  number={4},
  pages={253--350},
  year={1989},
  publisher={Taylor \& Francis},
  doi={https://doi.org/10.1207/s15327051hci0404_1}
}

@book{Graeber_2001, address={New York}, title={Toward An Anthropological Theory of Value}, rights={http://www.springer.com/tdm}, ISBN={9780312240455}, url={http://link.springer.com/10.1057/9780312299064}, DOI={10.1057/9780312299064}, publisher={Palgrave Macmillan US}, author={Graeber, David}, year={2001}, language={en} }

@book{Graeber_2019, address={New York}, edition={First Simon \&Schuster trade paperback edition}, title={Bullshit jobs}, ISBN={9781501143335}, callNumber={HF5549.5.J63 G73 2019}, publisher={Simon \& Schuster Paperbacks}, author={Graeber, David}, year={2019} }

@article{omidi2023labor,
  title={Labor process theory and critical HRM: A systematic review and agenda for future research},
  author={Omidi, Afshin and Dal Zotto, Cinzia and Gandini, Alessandro},
  journal={European Management Journal},
  volume={41},
  number={6},
  pages={899--913},
  year={2023},
  publisher={Elsevier},
  doi={https://doi.org/10.1016/j.emj.2023.05.003}
}

@article{Barry_1981, title={Book Reviews: In The Name Of Efficiency: Management Theory And Shopfloor Practice}, volume={13}, rights={https://journals.sagepub.com/page/policies/text-and-data-mining-license}, ISSN={0486-6134, 1552-8502}, url={https://journals.sagepub.com/doi/10.1177/048661348201300408}, DOI={10.1177/048661348201300408}, number={4}, journal={Review of Radical Political Economics}, author={Barry, Janis}, year={1981}, month=dec, pages={65–66}, language={en} }

@article{syverson2009determines,
  title={What determines productivity at the micro level?},
  author={Syverson, Chad and Goolsbee, Austan and Levitt, Steven},
  journal={Work. Pap., Univ. Chicago},
  year={2009}
}

@article{learned2009impact,
  title={Impact Assessments in Finance and Private Sector Development: What Have We Learned ?and What Should We Learn},
  author={David McKenzie},
  year={2010},
  publisher={Oxford University Press},
  url={https://www.jstor.org/stable/pdf/40891374.pdf}
}

@article{uddin2015evolution,
  title={Evolution of modern management through Taylorism: An adjustment of Scientific Management comprising behavioral science},
  author={Uddin, Nasir and Hossain, Fariha},
  journal={Procedia Computer Science},
  volume={62},
  pages={578--584},
  year={2015},
  publisher={Elsevier}
}

@inproceedings{cha25chi,
author = {Cha, Inha and Wong, Richmond Y.},
title = {Understanding Socio-technical Factors Configuring AI Non-Use in UX Work Practices},
year = {2025},
isbn = {9798400713941},
publisher = {Association for Computing Machinery},
address = {New York, NY, USA},
url = {https://doi.org/10.1145/3706598.3713140},
doi = {10.1145/3706598.3713140},
booktitle = {Proceedings of the 2025 CHI Conference on Human Factors in Computing Systems},
articleno = {1110},
numpages = {17},
location = {
},
series = {CHI '25}
}

@article{mariani2023types,
  title={Types of innovation and artificial intelligence: A systematic quantitative literature review and research agenda},
  author={Mariani, Marcello M and Machado, Isa and Nambisan, Satish},
  journal={Journal of Business Research},
  volume={155},
  pages={113364},
  year={2023},
  publisher={Elsevier}
}

@inproceedings{10.1145/3706598.3713479,
author = {Mei, Yihan and Wu, Zhao and Yu, Junnan and Li, Wenan and Zhou, Zhibin},
title = {GeneyMAP: Exploring the Potential of GenAI to Facilitate Mapping User Journeys for UX Design},
year = {2025},
isbn = {9798400713941},
publisher = {Association for Computing Machinery},
address = {New York, NY, USA},
url = {https://doi.org/10.1145/3706598.3713479},
doi = {10.1145/3706598.3713479},
booktitle = {Proceedings of the 2025 CHI Conference on Human Factors in Computing Systems},
articleno = {270},
numpages = {22},
location = {
},
series = {CHI '25}
}

@inproceedings{10.1145/3613904.3642847,
author = {Son, Kihoon and Choi, DaEun and Kim, Tae Soo and Kim, Young-Ho and Kim, Juho},
title = {GenQuery: Supporting Expressive Visual Search with Generative Models},
year = {2024},
isbn = {9798400703300},
publisher = {Association for Computing Machinery},
address = {New York, NY, USA},
url = {https://doi.org/10.1145/3613904.3642847},
doi = {10.1145/3613904.3642847},
booktitle = {Proceedings of the 2024 CHI Conference on Human Factors in Computing Systems},
articleno = {180},
numpages = {19},
location = {Honolulu, HI, USA},
series = {CHI '24}
}

@inproceedings{10.1145/3706599.3720189,
author = {Choi, DaEun and Son, Kihoon and Jung, HyunJoon and Kim, Juho},
title = {Expandora: Broadening Design Exploration with Text-to-Image Model},
year = {2025},
isbn = {9798400713958},
publisher = {Association for Computing Machinery},
address = {New York, NY, USA},
url = {https://doi.org/10.1145/3706599.3720189},
doi = {10.1145/3706599.3720189},
booktitle = {Proceedings of the Extended Abstracts of the CHI Conference on Human Factors in Computing Systems},
articleno = {232},
numpages = {10},
location = {
},
series = {CHI EA '25}
}

@inproceedings{10.1145/3613904.3642794,
author = {Choi, DaEun and Hong, Sumin and Park, Jeongeon and Chung, John Joon Young and Kim, Juho},
title = {CreativeConnect: Supporting Reference Recombination for Graphic Design Ideation with Generative AI},
year = {2024},
isbn = {9798400703300},
publisher = {Association for Computing Machinery},
address = {New York, NY, USA},
url = {https://doi.org/10.1145/3613904.3642794},
doi = {10.1145/3613904.3642794},
booktitle = {Proceedings of the 2024 CHI Conference on Human Factors in Computing Systems},
articleno = {1055},
numpages = {25},
location = {Honolulu, HI, USA},
series = {CHI '24}
}

@inproceedings{shin2025what,
    title = {What About My Design Context?: Exploring the Use of Generative AI to Support Customization of Translational Research Artifacts},
    author = {Shin, Donghoon and Chen, Alex and Hsieh, Gary and Wang, Lucy Lu},
    booktitle = {Proceedings of the 2025 ACM Designing Interactive Systems Conference},
    year = {2025},
    doi={https://doi.org/10.1145/3715336.3735686}
  }

@inproceedings{10.1145/3706598.3713273,
author = {Shin, Joongi and Polyanskaya, Anna and Lucero, Andr\'{e}s and Oulasvirta, Antti},
title = {No Evidence for LLMs Being Useful in Problem Reframing},
year = {2025},
isbn = {9798400713941},
publisher = {Association for Computing Machinery},
address = {New York, NY, USA},
url = {https://doi.org/10.1145/3706598.3713273},
doi = {10.1145/3706598.3713273},
booktitle = {Proceedings of the 2025 CHI Conference on Human Factors in Computing Systems},
articleno = {243},
numpages = {25},
location = {
},
series = {CHI '25}
}

@inproceedings{liCHI24,
author = {Li, Jie and Cao, Hancheng and Lin, Laura and Hou, Youyang and Zhu, Ruihao and El Ali, Abdallah},
title = {User Experience Design Professionals’ Perceptions of Generative Artificial Intelligence},
year = {2024},
isbn = {9798400703300},
publisher = {Association for Computing Machinery},
address = {New York, NY, USA},
url = {https://doi.org/10.1145/3613904.3642114},
doi = {10.1145/3613904.3642114},
booktitle = {Proceedings of the CHI Conference on Human Factors in Computing Systems},
articleno = {381},
numpages = {18},
location = {Honolulu, HI, USA},
series = {CHI '24}
}

@incollection{li2021ml,
  title={ML Tools for the Web: A Way for Rapid Prototyping and HCI Research},
  author={Li, Na and Mayes, Jason and Yu, Ping},
  booktitle={Artificial Intelligence for Human Computer Interaction: A Modern Approach},
  pages={315--343},
  year={2021},
  publisher={Springer}
}

@inproceedings{10.1145/3643834.3660703,
author = {Wang, Ziyan and Shen, Luyao and Kuang, Emily and Zhang, Shumeng and Fan, Mingming},
title = {Exploring the Impact of Artificial Intelligence-Generated Content (AIGC) Tools on Social Dynamics in UX Collaboration},
year = {2024},
isbn = {9798400705830},
publisher = {Association for Computing Machinery},
address = {New York, NY, USA},
url = {https://doi.org/10.1145/3643834.3660703},
doi = {10.1145/3643834.3660703},
booktitle = {Proceedings of the 2024 ACM Designing Interactive Systems Conference},
pages = {1594–1606},
numpages = {13},
location = {Copenhagen, Denmark},
series = {DIS '24}
}

@inproceedings{10.1145/1357054.1357156,
author = {Wright, Peter and McCarthy, John},
title = {Empathy and experience in HCI},
year = {2008},
isbn = {9781605580111},
publisher = {Association for Computing Machinery},
address = {New York, NY, USA},
url = {https://doi.org/10.1145/1357054.1357156},
doi = {10.1145/1357054.1357156},
booktitle = {Proceedings of the SIGCHI Conference on Human Factors in Computing Systems},
pages = {637–646},
numpages = {10},
location = {Florence, Italy},
series = {CHI '08}
}

@book{greenbaum1979name,
  title={In the name of efficiency: Management theory and shopfloor practice in data processing work},
  author={Greenbaum, Joan M},
  year={1979},
  publisher={Temple University}
}

@article{dix2017human,
  title={Human--computer interaction, foundations and new paradigms},
  author={Dix, Alan},
  journal={Journal of Visual Languages \& Computing},
  volume={42},
  pages={122--134},
  year={2017},
  publisher={Elsevier}
}

@book{blythe2004funology,
  title={Funology: from usability to enjoyment},
  author={Blythe, Mark A and Overbeeke, Kees and Monk, Andrew F and Wright, Peter C},
  year={2004},
  publisher={Springer},
}

@article{carroll2001evolution,
  title={The evolution of human-computer interaction},
  author={Carroll, John M},
  journal={Annual Review of Psychology},
  volume={48},
  pages={501--522},
  year={2001},
  url={https://ieeexplore.ieee.org/document/9263261/}
}

@article{myers1998brief,
  title={A brief history of human-computer interaction technology},
  author={Myers, Brad A},
  journal={interactions},
  volume={5},
  number={2},
  pages={44--54},
  year={1998},
  publisher={ACM New York, NY, USA}
}

@article{waterson2011world,
  title={World War II and other historical influences on the formation of the Ergonomics Research Society},
  author={Waterson, Patrick},
  journal={Ergonomics},
  volume={54},
  number={12},
  pages={1111--1129},
  year={2011},
  publisher={Taylor \& Francis}
}

@article{grudin2005three,
  title={Three faces of human-computer interaction},
  author={Grudin, Jonathan},
  journal={IEEE Annals of the History of Computing},
  volume={27},
  number={4},
  pages={46--62},
  year={2005},
  publisher={IEEE}
}

@inproceedings{shin24DIS,
author = {Shin, Joongi and Hedderich, Michael A. and Rey, Bart\l{}omiej Jakub and Lucero, Andr\'{e}s and Oulasvirta, Antti},
title = {Understanding Human-AI Workflows for Generating Personas},
year = {2024},
isbn = {9798400705830},
publisher = {Association for Computing Machinery},
address = {New York, NY, USA},
url = {https://doi.org/10.1145/3643834.3660729},
doi = {10.1145/3643834.3660729},
booktitle = {Proceedings of the 2024 ACM Designing Interactive Systems Conference},
pages = {757–781},
numpages = {25},
location = {Copenhagen, Denmark},
series = {DIS '24}
}

@inproceedings{10.1145/3715275.3732198,
author = {Tseng, Emily and Young, Meg and Le Qu\'{e}r\'{e}, Marianne Aubin and Rinehart, Aimee and Suresh, Harini},
title = {"Ownership, Not Just Happy Talk": Co-Designing a Participatory Large Language Model for Journalism},
year = {2025},
isbn = {9798400714825},
publisher = {Association for Computing Machinery},
address = {New York, NY, USA},
url = {https://doi.org/10.1145/3715275.3732198},
doi = {10.1145/3715275.3732198},
booktitle = {Proceedings of the 2025 ACM Conference on Fairness, Accountability, and Transparency},
pages = {3119–3130},
numpages = {12},
location = {
},
series = {FAccT '25}
}

@inproceedings{ding2025considering,
  title={Considering Large Language Model Integration in Expressive Computer Science Learning Environments for Blind and Visually Impaired Learners Through Co-design},
  author={Ding, Shi and Smith, Jason Brent and Magerko, Brian},
  booktitle={International Conference on Artificial Intelligence in Education},
  pages={472--480},
  year={2025},
  organization={Springer},
  doi={https://doi.org/10.1007/978-3-031-98459-4_36}
}

@inproceedings{10.1145/3377325.3377522,
author = {Karimi, Pegah and Rezwana, Jeba and Siddiqui, Safat and Maher, Mary Lou and Dehbozorgi, Nasrin},
title = {Creative sketching partner: an analysis of human-AI co-creativity},
year = {2020},
isbn = {9781450371186},
publisher = {Association for Computing Machinery},
address = {New York, NY, USA},
url = {https://doi.org/10.1145/3377325.3377522},
doi = {10.1145/3377325.3377522},
booktitle = {Proceedings of the 25th International Conference on Intelligent User Interfaces},
pages = {221–230},
numpages = {10},
location = {Cagliari, Italy},
series = {IUI '20}
}

@inproceedings{10.1145/3643834.3660720,
author = {Takaffoli, Macy and Li, Sijia and M\"{a}kel\"{a}, Ville},
title = {Generative AI in User Experience Design and Research: How Do UX Practitioners, Teams, and Companies Use GenAI in Industry?},
year = {2024},
isbn = {9798400705830},
publisher = {Association for Computing Machinery},
address = {New York, NY, USA},
url = {https://doi.org/10.1145/3643834.3660720},
doi = {10.1145/3643834.3660720},
booktitle = {Proceedings of the 2024 ACM Designing Interactive Systems Conference},
pages = {1579–1593},
numpages = {15},
location = {Copenhagen, Denmark},
series = {DIS '24}
}

@article{dwivedi2019re,
  title={Re-examining the unified theory of acceptance and use of technology (UTAUT): Towards a revised theoretical model},
  author={Dwivedi, Yogesh K and Rana, Nripendra P and Jeyaraj, Anand and Clement, Marc and Williams, Michael D},
  journal={Information systems frontiers},
  volume={21},
  number={3},
  pages={719--734},
  year={2019},
  publisher={Springer}
}

@inproceedings{10.1145/3706598.3713978,
author = {Gomez-Beldarrain, Garoa and Verma, Himanshu and Kim, Euiyoung and Bozzon, Alessandro},
title = {Why does Automation Adoption in Organizations Remain a Fallacy?: Scrutinizing Practitioners' Imaginaries in an International Airport},
year = {2025},
isbn = {9798400713941},
publisher = {Association for Computing Machinery},
address = {New York, NY, USA},
url = {https://doi.org/10.1145/3706598.3713978},
doi = {10.1145/3706598.3713978},
booktitle = {Proceedings of the 2025 CHI Conference on Human Factors in Computing Systems},
articleno = {213},
numpages = {19},
location = {
},
series = {CHI '25}
}

@misc{WGAAI,
  author       = {{Writers Guild of America West}},
  title        = {Artificial Intelligence — Know Your Rights},
  howpublished = {\url{https://www.wga.org/contracts/know-your-rights/artificial-intelligence}},
  year         = {2023},
  note         = {Accessed: 2025-09-01},
}

@inproceedings{10.1145/3744169.3744170,
author = {Baumer, Eric P. S. and Khovanskaya, Vera},
title = {Sociotechnical Remantling},
year = {2025},
isbn = {9798400720031},
publisher = {Association for Computing Machinery},
address = {New York, NY, USA},
url = {https://doi.org/10.1145/3744169.3744170},
doi = {10.1145/3744169.3744170},
booktitle = {Proceedings of the Sixth Decennial Aarhus Conference: Computing X Crisis},
pages = {27–41},
numpages = {15},
location = {
},
series = {AAR '25}
}

@article{verganti2020innovation,
  title={Innovation and design in the age of artificial intelligence},
  author={Verganti, Roberto and Vendraminelli, Luca and Iansiti, Marco},
  journal={Journal of product innovation management},
  volume={37},
  number={3},
  pages={212--227},
  year={2020},
  publisher={Wiley Online Library}
}

@inproceedings{10.1145/3025453.3025742,
author = {Lindley, Joseph and Coulton, Paul and Sturdee, Miriam},
title = {Implications for Adoption},
year = {2017},
isbn = {9781450346559},
publisher = {Association for Computing Machinery},
address = {New York, NY, USA},
url = {https://doi.org/10.1145/3025453.3025742},
doi = {10.1145/3025453.3025742},
booktitle = {Proceedings of the 2017 CHI Conference on Human Factors in Computing Systems},
pages = {265–277},
numpages = {13},
location = {Denver, Colorado, USA},
series = {CHI '17}
}

@article{davis1989technology,
  title={Technology acceptance model: TAM},
  author={Davis, Fred D and others},
  journal={Al-Suqri, MN, Al-Aufi, AS: Information Seeking Behavior and Technology Adoption},
  volume={205},
  number={219},
  pages={5},
  year={1989}
}

@article{hassenzahl2006user,
  title={User experience-a research agenda},
  author={Hassenzahl, Marc and Tractinsky, Noam},
  journal={Behaviour \& information technology},
  volume={25},
  number={2},
  pages={91--97},
  year={2006},
  publisher={Taylor \& Francis}
}

@article{hassenzahl2004interplay,
  title={The interplay of beauty, goodness, and usability in interactive products},
  author={Hassenzahl, Marc},
  journal={Human--Computer Interaction},
  volume={19},
  number={4},
  pages={319--349},
  year={2004},
  publisher={Taylor \& Francis}
}

@article{o2010influence,
  title={The influence of hedonic and utilitarian motivations on user engagement: The case of online shopping experiences},
  author={O’Brien, Heather Lynn},
  journal={Interacting with computers},
  volume={22},
  number={5},
  pages={344--352},
  year={2010},
  publisher={Oxford University Press Oxford, UK}
}

@article{10.1145/3127358,
author = {Hornb\ae{}k, Kasper and Hertzum, Morten},
title = {Technology Acceptance and User Experience: A Review of the Experiential Component in HCI},
year = {2017},
issue_date = {October 2017},
publisher = {Association for Computing Machinery},
address = {New York, NY, USA},
volume = {24},
number = {5},
issn = {1073-0516},
url = {https://doi.org/10.1145/3127358},
doi = {10.1145/3127358},
journal = {ACM Trans. Comput.-Hum. Interact.},
month = oct,
articleno = {33},
numpages = {30}
}

@inproceedings{10.1145/2702613.2724724,
author = {Chilana, Parmit K. and Czerwinski, Mary P. and Grossman, Tovi and Harrison, Chris and Kumar, Ranjitha and Parikh, Tapan S. and Zhai, Shumin},
title = {Technology Transfer of HCI Research Innovations: Challenges and Opportunities},
year = {2015},
isbn = {9781450331463},
publisher = {Association for Computing Machinery},
address = {New York, NY, USA},
url = {https://doi.org/10.1145/2702613.2724724},
doi = {10.1145/2702613.2724724},
booktitle = {Proceedings of the 33rd Annual ACM Conference Extended Abstracts on Human Factors in Computing Systems},
pages = {823–828},
numpages = {6},
location = {Seoul, Republic of Korea},
series = {CHI EA '15}
}

@article{cooper1995representing,
  title={Representing the user: Notes on the disciplinary rhetoric of human-computer},
  author={Cooper, Geoff and Bowers, John},
  journal={The social and interactional dimensions of human-computer interfaces},
  volume={48},
  year={1995},
  publisher={Cambridge University Press}
}

@misc{yee2025superagency,
  title={Superagency in the workplace: Empowering people to unlock AI’s full potential. McKinsey \& Company Report, January 28},
  author={Yee, L and Chui, M and Roberts, R},
  year={2025}
}

@article{kassa2025impact,
  title={The impact of artificial intelligence on organizational performance: The mediating role of employee productivity},
  author={Kassa, Belayneh Yitayew and Worku, Eyob Ketema},
  journal={Journal of open innovation: technology, market, and complexity},
  volume={11},
  number={1},
  pages={100474},
  year={2025},
  publisher={Elsevier},
  doi={https://doi.org/10.1016/j.joitmc.2025.100474}
}

@inproceedings{10.1145/240080.240259,
author = {Greenbaum, Joan},
title = {Back to labor: returning to labor process discussions in the study of work},
year = {1996},
isbn = {0897917650},
publisher = {Association for Computing Machinery},
address = {New York, NY, USA},
url = {https://doi.org/10.1145/240080.240259},
doi = {10.1145/240080.240259},
booktitle = {Proceedings of the 1996 ACM Conference on Computer Supported Cooperative Work},
pages = {229–237},
numpages = {9},
location = {Boston, Massachusetts, USA},
series = {CSCW '96}
}

@inproceedings{10.1145/3491101.3516386,
author = {Ahmed, Alex A.},
title = {Who Owns the Future of Work?},
year = {2022},
isbn = {9781450391566},
publisher = {Association for Computing Machinery},
address = {New York, NY, USA},
url = {https://doi.org/10.1145/3491101.3516386},
doi = {10.1145/3491101.3516386},
booktitle = {Extended Abstracts of the 2022 CHI Conference on Human Factors in Computing Systems},
articleno = {3},
numpages = {6},
location = {New Orleans, LA, USA},
series = {CHI EA '22}
}

@article{10.1145/3641001,
author = {Jing, Felicia S. and Berger, Sara E. and Becerra Sandoval, Juana Catalina and Pepper, Kristin and Wheeler, April M. and Mayoral, Paula Redondo and Lokesh, Divya and Feng, Alice and Mijalkovic, Marija and Bao, Chaoyun and Dholakia, Sara and Goyal, Mohit},
title = {Designing for Agonism: 12 Workers' Perspectives on Contesting Technology Futures},
year = {2024},
issue_date = {April 2024},
publisher = {Association for Computing Machinery},
address = {New York, NY, USA},
volume = {8},
number = {CSCW1},
url = {https://doi.org/10.1145/3641001},
doi = {10.1145/3641001},
journal = {Proc. ACM Hum.-Comput. Interact.},
month = apr,
articleno = {162},
numpages = {25}
}

@inproceedings{10.1145/3715070.3747335,
author = {Sum, Cella M.},
title = {From the Future of Work to the Future of Labor: Centering Worker Resistance in the Age of AI and Automation},
year = {2025},
isbn = {9798400714801},
publisher = {Association for Computing Machinery},
address = {New York, NY, USA},
url = {https://doi.org/10.1145/3715070.3747335},
doi = {10.1145/3715070.3747335},
booktitle = {Companion Publication of the 2025 Conference on Computer-Supported Cooperative Work and Social Computing},
pages = {19–21},
numpages = {3},
location = {
},
series = {CSCW Companion '25}
}

@inproceedings{10.1145/3744169.3744179,
author = {Lin, Cindy Kaiying and Dombrowski, Lynn and Bardzell, Shaowen},
title = {Whose, Which, and What Crisis? A Critical Analysis of Crisis in Computing Supply Chains},
year = {2025},
isbn = {9798400720031},
publisher = {Association for Computing Machinery},
address = {New York, NY, USA},
url = {https://doi.org/10.1145/3744169.3744179},
doi = {10.1145/3744169.3744179},
booktitle = {Proceedings of the Sixth Decennial Aarhus Conference: Computing X Crisis},
pages = {56–70},
numpages = {15},
location = {
},
series = {AAR '25}
}

@article{10.1145/3686899,
author = {Toxtli, Carlos and Curtis, Christopher and Savage, Saiph},
title = {A Culturally-Aware AI Tool for Crowdworkers: Leveraging Chronemics to Support Diverse Work Styles},
year = {2024},
issue_date = {November 2024},
publisher = {Association for Computing Machinery},
address = {New York, NY, USA},
volume = {8},
number = {CSCW2},
url = {https://doi.org/10.1145/3686899},
doi = {10.1145/3686899},
journal = {Proc. ACM Hum.-Comput. Interact.},
month = nov,
articleno = {360},
numpages = {34}
}

@article{beaudry2005changes,
  title={Changes in US wages, 1976--2000: ongoing skill bias or major technological change?},
  author={Beaudry, Paul and Green, David A},
  journal={Journal of Labor Economics},
  volume={23},
  number={3},
  pages={609--648},
  year={2005},
  url={https://economics.ubc.ca/wp-content/uploads/sites/38/2013/05/pdf_paper_david-green-changes-us-wages-1976-2000.pdf},
  publisher={The University of Chicago Press}
}

@inproceedings{10.1145/3715070.3748296,
author = {Lushnikova, Alina and Muller, Michael and Bardzell, Shaowen and Li, Toby Jia-Jun and Savage, Saiph},
title = {CSCW Contributions to Critical Futures of Work},
year = {2025},
isbn = {9798400714801},
publisher = {Association for Computing Machinery},
address = {New York, NY, USA},
url = {https://doi.org/10.1145/3715070.3748296},
doi = {10.1145/3715070.3748296},
booktitle = {Companion Publication of the 2025 Conference on Computer-Supported Cooperative Work and Social Computing},
pages = {161–167},
numpages = {7},
location = {
},
series = {CSCW Companion '25}
}

@inproceedings{10.1145/3613904.3642902,
author = {Tankelevitch, Lev and Kewenig, Viktor and Simkute, Auste and Scott, Ava Elizabeth and Sarkar, Advait and Sellen, Abigail and Rintel, Sean},
title = {The Metacognitive Demands and Opportunities of Generative AI},
year = {2024},
isbn = {9798400703300},
publisher = {Association for Computing Machinery},
address = {New York, NY, USA},
url = {https://doi.org/10.1145/3613904.3642902},
doi = {10.1145/3613904.3642902},
booktitle = {Proceedings of the 2024 CHI Conference on Human Factors in Computing Systems},
articleno = {680},
numpages = {24},
location = {Honolulu, HI, USA},
series = {CHI '24}
}

@article{spanjol2018successive,
  title={Successive innovation in digital and physical products: Synthesis, conceptual framework, and research directions},
  author={Spanjol, Jelena and Xiao, Yazhen and Welzenbach, Lisa},
  journal={Innovation and Strategy},
  volume={15},
  pages={31--62},
  year={2018},
  publisher={Emerald Publishing Limited}
}

@inproceedings{10.1145/2702123.2702412,
author = {Chilana, Parmit K. and Ko, Amy J. and Wobbrock, Jacob},
title = {From User-Centered to Adoption-Centered Design: A Case Study of an HCI Research Innovation Becoming a Product},
year = {2015},
isbn = {9781450331456},
publisher = {Association for Computing Machinery},
address = {New York, NY, USA},
url = {https://doi.org/10.1145/2702123.2702412},
doi = {10.1145/2702123.2702412},
booktitle = {Proceedings of the 33rd Annual ACM Conference on Human Factors in Computing Systems},
pages = {1749–1758},
numpages = {10},
location = {Seoul, Republic of Korea},
series = {CHI '15}
}

@inproceedings{10.1145/3698061.3726924,
author = {Rhys Cox, Samuel and B\o{}jer Djern\ae{}s, Helena and van Berkel, Niels},
title = {Beyond Productivity: Rethinking the Impact of Creativity Support Tools},
year = {2025},
isbn = {9798400712890},
publisher = {Association for Computing Machinery},
address = {New York, NY, USA},
url = {https://doi.org/10.1145/3698061.3726924},
doi = {10.1145/3698061.3726924},
booktitle = {Proceedings of the 2025 Conference on Creativity and Cognition},
pages = {735–749},
numpages = {15},
location = {
},
series = {C\&C '25}
}

@incollection{rogers2014diffusion,
  title={Diffusion of innovations},
  author={Rogers, Everett M and Singhal, Arvind and Quinlan, Margaret M},
  booktitle={An integrated approach to communication theory and research},
  pages={432--448},
  year={2014},
  publisher={Routledge}
}

@incollection{miller2015rogers,
  title={Rogers' innovation diffusion theory (1962, 1995)},
  author={Miller, Rebecca L},
  booktitle={Information seeking behavior and technology adoption: Theories and trends},
  pages={261--274},
  year={2015},
  publisher={IGI Global Scientific Publishing}
}

@inproceedings{wieczorek2025architecting,
  title={Architecting Utopias: How AI in Healthcare Envisions Societal Ideals and Human Flourishing},
  author={Wieczorek, Catherine and Biggs, Heidi and Payyapilly Thiruvenkatanathan, Kamala and Bardzell, Shaowen},
  booktitle={Proceedings of the 2025 CHI Conference on Human Factors in Computing Systems},
  pages={1--15},
  year={2025},
  doi={https://doi.org/10.1145/3706598.3713118}
}

@book{d2023data,
  title={Data feminism},
  author={D'ignazio, Catherine and Klein, Lauren F},
  year={2023},
  publisher={MIT press}
}

@book{rosner2018critical,
  title={Critical fabulations: Reworking the methods and margins of design},
  author={Rosner, Daniela K},
  year={2018},
  publisher={MIT press}
}

@inproceedings{irani2010postcolonial,
  title={Postcolonial computing: a lens on design and development},
  author={Irani, Lilly and Vertesi, Janet and Dourish, Paul and Philip, Kavita and Grinter, Rebecca E},
  booktitle={Proceedings of the SIGCHI conference on human factors in computing systems},
  pages={1311--1320},
  year={2010},
  doi={https://doi.org/10.1145/1753326.1753522}
}

@inproceedings{karusala2023unsettling,
  title={Unsettling care infrastructures: from the individual to the structural in a digital maternal and child health intervention},
  author={Karusala, Naveena and G, Victoria and Yan, Shirley and Anderson, Richard},
  booktitle={Proceedings of the 2023 CHI Conference on Human Factors in Computing Systems},
  pages={1--16},
  year={2023},
  doi={https://doi.org/10.1145/3544548.3581553}
}

\appendix
\section{Workshop Materials}
\label{appendix:materials}

\begin{figure}
    \centering
    \includegraphics[width=0.9\linewidth]{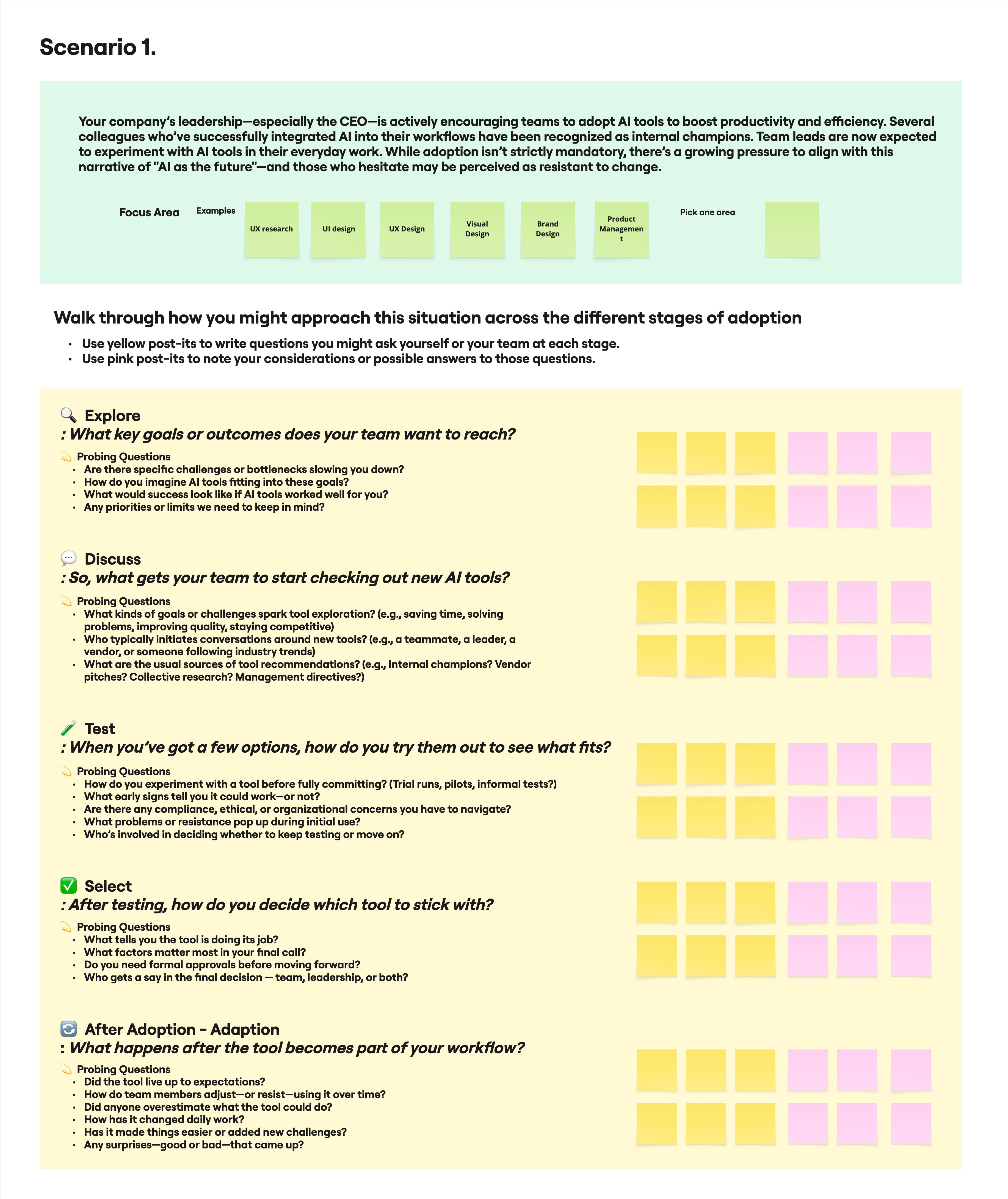}
    \caption{A Screenshot of Workshop Miro board}
    \label{fig:miro}
\end{figure}

\subsubsection{Scenarios}

During the workshop, we provide three scenarios that situate AI adoption challenges in distinct contexts:

\begin{itemize}
    \item Top-down pressure: Company leadership urges teams to adopt AI tools for productivity, creating tension between organizational expectations and team readiness.

    \item Regulated domain: A healthcare client requires strict compliance with laws such as HIPAA, foregrounding issues of trust, privacy, and regulatory adherence.

    \item Team conflict: A design team debates whether to adopt generative AI, highlighting enthusiasm, skepticism, and the potential impact on collaboration.

\end{itemize}

\subsubsection{The situation across the different stages of adoption}

Each scenario was paired with a structured framework that guided participants through five sequential stages of AI adoption, supported by probing questions to surface considerations, challenges, and decision-making practices. The \textit{Discuss} stage centered on identifying the problem or opportunity that initiated interest in AI, prompting participants to reflect on who raised the issue, why it mattered, and what goals motivated exploration. In the \textit{Explore} stage, participants considered how they or their teams might learn about available tools, which features or criteria would guide their choices, and what concerns could surface early on. The \textit{Test} stage focused on envisioning how tools would be trialed, including who would participate, how effectiveness might be assessed, and what forms of feedback would be most valuable. During the \textit{Select} stage, participants examined how final adoption decisions are made, what criteria and trade-offs carry weight, and whose authority shapes outcomes. Finally, in the \textit{Adapt} stage, they reflected on the long-term integration of chosen tools, considering how workflows might shift, how teams might adjust or resist, and what new challenges or unintended consequences could arise. Importantly, we emphasized that participants were free to reorder or reinterpret these stages to fit their own organizational contexts.

We guided participants worked with color-coded post-its, with yellow used for recording questions participants have regarding this stage and pink used for reflections or considerations in the stage. This format encouraged brainstorming and layered responses across the different adoption stages.

\section{Example Follow-up Interview Protocol}
\label{appendix:interviewguide}

\subsection*{1. Participant Background and Working Style}
\begin{itemize}
    \item Could you briefly describe your role and what a typical day looks like?
    \item How is the UX or research function structured in your organization?
    \item How many designers are in your organization, and how are they distributed across teams or products?
    \item Are you part of a centralized design/research team, or embedded within product or feature teams?
    \item How do you communicate and collaborate with designers across different teams?
    \item How are design tools or conventions selected, and who is responsible for these decisions?
    \item What are the primary priorities or goals when you design a feature or service?
\end{itemize}

\subsection*{2. Walk-through of Actual AI Adoption}
\begin{itemize}
    \item Where do you currently use AI tools in your workflow?
    \item Can you walk through a recent example of adopting or trying an AI tool?
    \item How does your organization evaluate or approve new tools?
    \item In your own work, how do you decide whether to use or not to use an AI tool?
    \item What criteria matter most when evaluating new tools?
\end{itemize}

\subsection*{3. Reflections on the Adoption Process}
\begin{itemize}
    \item What challenges or trade-offs come up when adopting AI tools?
    \item What emotions or concerns emerge in this process?
    \item Have AI-supported processes changed how your work feels creatively, socially, or emotionally?
    \item Do AI-driven workflows feel experimental, or more like established practice?
    \item Are there organizational, social, or ethical concerns that come up? Can you share examples?
    \item Is there anything about the pace or logic of AI-supported work that feels misaligned with how you prefer to design or collaborate?
    \item How do you define or recognize successful tool adoption?
\end{itemize}

\subsection*{4. Feedback on the Workshop Experience}
\begin{itemize}
    \item Did the workshop prompt any new reflections?
    \item Which parts of the activities or toolkit were most useful? Which could be improved?
    \item If these questions or activities were used at your workplace, when in the process would they be most helpful?
    \item Do you have time and space for this kind of deliberation? If not, what are the barriers?
    \item If the workshop were conducted asynchronously, how should it be structured?
\end{itemize}

\subsection*{5. Broader Reflections on AI in Practice}
\begin{itemize}
    \item Before this study, had you ever thought deeply about AI adoption in your work?
    \item When you think about the growing role of AI in your industry, what feelings come up?
    \item Who decides which AI tools are adopted in your workplace? How involved are you in those decisions?
    
\end{itemize}

\subsection*{6. Future of UX Work in Relation to AI}
\begin{itemize}
    \item How might AI shape the future of UX roles or responsibilities?
    \item What forms of participation should designers or practitioners have in AI tool decisions?
    \item If someone from your organization could have joined this workshop, who would that be and why?
    \item What kinds of support or resources would help teams make more thoughtful decisions about AI adoption?
\end{itemize}

%%%ack 
%%sarah
%%%%%pooja
%%Sylvia

\end{document}